\documentclass[12pt]{article}
\usepackage{graphicx}
\usepackage{cite}
\begin{document}

\def \d {{\rm d}}
\def \D {{\rm D}}
\def \e {{\rm e}}
\def \iu {{\rm i}\,}

\newcommand{\boldpartial}{\mbox{\boldmath${\partial}$}} 
\newcommand{\bolda}{\mbox{\boldmath$a$}} 
\newcommand{\boldd}{\textbf{d}} 
\newcommand{\boldomega}{\mbox{\boldmath$\omega$}} 
\newcommand{\Boldomega}{\mbox{\boldmath$\Omega$}} 
\newcommand{\boldu}{\mbox{\boldmath$u$}} 
\newcommand{\bolde}{\mbox{\boldmath$e$}} 
\newcommand{\boldv}{\mbox{\boldmath$v$}} 
\newcommand{\boldk}{\mbox{\boldmath$k$}} 
\newcommand{\boldl}{\mbox{\boldmath$l$}} 
\newcommand{\boldm}{\mbox{\boldmath$m$}} 
\newcommand{\bboldm}{\bar{\mbox{\boldmath$m$}}} 
\newcommand{\boldZ}{\mbox{\boldmath$Z$}} 

\newcommand{\ssqrt}{{\textstyle\frac1{\sqrt{2}}}} 

\title{Gyratonic \emph{pp}-waves and their impulsive limit}

\author{
J.~Podolsk\'y$^1$\thanks{{\tt podolsky@mbox.troja.mff.cuni.cz}},
R.~Steinbauer$^2$\thanks{{\tt roland.steinbauer@univie.ac.at}}
and R.~\v{S}varc$^1$\thanks{{\tt robert.svarc@mff.cuni.cz}} \\ \\
$^1$ Institute of Theoretical Physics,\\
Charles University in Prague, Faculty of Mathematics and Physics, \\
V Hole\v{s}ovi\v{c}k\'ach 2, 18000 Prague 8, Czech Republic.\\ \\
$^2$ Faculty of Mathematics, University of Vienna, \\
Oskar-Morgenstern-Platz 1, 1090 Vienna, Austria. \\ \\
}

\maketitle

\baselineskip=19pt

\begin{abstract}
We investigate a class of gravitational \emph{pp}-waves which
represent the exterior vacuum field of spinning
particles moving with the speed of light. Such exact spacetimes are described by
the original Brinkmann form of the \emph{pp}-wave
metric including the often neglected off-diagonal terms. We put emphasis on a
clear physical and geometrical interpretation
of these off-diagonal metric components. We explicitly analyze several new
properties of these spacetimes associated with the spinning character of the
source, such as rotational dragging of frames, geodesic deviation, impulsive limits and the corresponding behavior of geodesics.
\end{abstract}

\vfil\noindent
PACS class 04.20.Jb, 04.30.-w, 04.30.Nk, 04.30.Db

\bigskip\noindent
MSC class 83C15, 83C35

\bigskip\noindent
Keywords: Impulsive gravitational waves, \emph{pp}-waves, gyratons.


\vfil
\eject

\section{Introduction}
\label{intro}

In the present work we will study mathematical and physical properties of the family of spacetimes described by the (so called) \emph{pp}-wave metric
\begin{equation}\label{pp_metric}
\d s^2 = \delta_{ij}\,\d x^i\,\d x^j-2\,\d u\,\d r +2\,a_i(u,x^j)\,\d u\,\d x^i +H(u,x^j)\,\d u^2 \,,
\end{equation}
which was introduced by Brinkmann in 1925 \cite{Brinkmann:1925}. It is now a well-known fact that these \emph{pp}-waves belong to the larger Kundt family of spacetimes \cite{Stephani:2003,GriffithsPodolsky:2009} which admit a nontwisting, nonshearing and nonexpanding geodesic null congruence generated by the vector field~$\boldpartial_r$, the coordinate ${r \in (-\infty,\infty)}$ being the corresponding affine parameter. For \emph{pp}-waves such a vector field (representing a repeated principal null direction of the Weyl tensor) is covariantly constant, and all the metric functions are independent of $r$. Moreover, since for the metric (\ref{pp_metric}) the transverse Riemannian space spanned by the spatial coordinates $x^i$ on each wave-surface ${u=\,}$const. is flat, these \emph{pp}-waves also belong to the important class of VSI spacetimes for which all (polynomial) curvature scalar invariants vanish \cite{PravdaPravdovaColeyMilson:2002}. In fact, those metrics (\ref{pp_metric}) that are Ricci-flat are ``universal spacetimes'' in the sense that they solve the vacuum field equations of all gravitational theories with Lagrangian constructed from the metric, the Riemann tensor and its derivatives of arbitrary order \cite{ColeyGibbonsHervikPope:2008,HervikPravdaPravdova:2013}, for example of quadratic gravity.

Although the family of \emph{pp}-wave spacetimes has been thoroughly studied for many decades and became a ``textbook'' prototype of exact gravitational waves in Einstein's general relativity (and its various extensions), there still remain some interesting aspects of the metric (\ref{pp_metric}) which deserve attention. In particular, here we concentrate on the physical interpretation and consequences of the off-diagonal metric functions $a_i(u,x^j)$, where ${\,i,j=2,3\,}$ in four-dimensional spacetimes.\footnote{The \emph{pp}-wave metric (\ref{pp_metric}) has a natural extension to any higher number of spacetime dimensions $D$ by taking ${i,j=2,3,\ldots,D-1}$, in which case the transverse flat space is ${(D-2)}$-dimensional.} In vacuum regions it is a standard and common procedure to completely remove these functions by a gauge (coordinate) transformation. However, such a freedom is generally \emph{only local} and completely ignores the global (topological) properties of the spacetimes. By neglecting the metric functions $a_i(u,x^j)$ in (\ref{pp_metric}), an important physical attribute of the spacetime is eliminated, namely the possible rotational character of the source of the gravitational waves---its internal spin/helicity.

This interesting fact was first noticed in 1970 by Bonnor \cite{Bonnor:1970b,Griffiths:1972} who studied both the interior and the exterior field of a ``spinning null fluid'' in the class of axially symmetric \emph{pp}-wave spacetimes (see section 18.5 of \cite{GriffithsPodolsky:2009} for a review).
In the interior region the energy-momentum tensor was phenomenologically described by the radiation density $T{_{uu}=\varrho}$ and by the components ${T_{ui}=j_i}$ representing the spinning character of the source, encoded in the corresponding angular momentum. Spacetimes with such localized spinning sources, which are moving at the speed of light, were independently rediscovered in 2005 by Frolov and his collaborators who emphasized their possible physical application as a model of a particle and thus called them ``gyratons''. These \emph{pp}-wave-type gyratons were subsequently investigated in greater detail and also generalized to higher dimensions, supergravity, and various nonflat backgrounds in a wider Kundt class which may also include a cosmological constant or an additional electromagnetic field \cite{FrolovFursaev:2005,FrolovIsraelZelnikov:2005,FrolovZelnikov:2005,FrolovZelnikov:2006,FrolovLin:2006,CaldarelliKlemmZorzan:2007,YoshinoZelnikovFrolov:2007,KadlecovaZelnikovKrtousPodolsky:2009,
KadlecovaKrtous:2010,KrtousPodolskyZelnikovKadlecova:2012}.

In this contribution we complement and extend the previous studies by
explicitly investigating
various physical and mathematical properties of these spacetimes
which have not been looked at before. Thereby we put our emphasis on a clear
geometrical and physical interpretation of the off-diagonal metric functions and
various specific effects associated with the gyrating nature of the source.

However, our first aim is to give a compact review of the topic
using a unified formalism. In particular, after presenting the metric and the
curvature quantities in section~\ref{metricforms}, we completely integrate the
field equations in the vacuum region in section~\ref{integrating},
\emph{before} turning to the delicate point of gauge issues in
subsection~\ref{gaugefreedom}. To approach the topic in this order allows us to
uncover the physical and geometrical meaning of all the integration functions
introduced in section~\ref{integrating}.
Indeed, the whole section~\ref{interpretation} is
devoted to an in-depth analysis and interpretation of the properties of
the spacetimes (\ref{pp_metric}). After securing the fact that in general it is
\emph{necessary} to keep the off-diagonal terms in the metric
(subsection~\ref{gaugefreedom}) we use these metric functions in
subsection~\ref{energyANDmomentum} to express the relevant physical parameters
of the spacetimes---energy and angular momentum---in a transparent way. In
subsection~\ref{interpretationframe} we introduce a natural orthonormal
interpretation frame for any geodesic observer and its associated null frame.
After studying their behavior under gauge transformations we employ these frames
in subsection~\ref{rotation} to analyze the dragging effect exerted on the
spacetime by the gyratonic source. We derive the Newman--Penrose field scalars in
subsection~\ref{NPscalars} and determine the Petrov type of the spacetimes.
In section~\ref{frame} we further employ the interpretation frame to analyze the
geodesic deviation in an invariant manner. We explicitly derive the two polarization
wave-amplitudes describing the relative motion of test particles, and in
subsection~\ref{ASsolution} we specialize to
the case of the simplest gyraton which is the axially symmetric one constructed
from the Aichelburg--Sexl
solution in \cite{FrolovFursaev:2005}. In section~\ref{geodesics} we briefly
analyze geodesic motion in
general gyratonic \emph{pp}-waves before turning to a deeper discussion of
impulsive limits in this class of spacetimes
(section~\ref{impulsive limit}). Here we resolve the delicate matter of the
possible coupling of the energy and the angular momentum density profiles.
Finally, in section~\ref{deodinimpuls} we discuss the geodescic equation in
impulsive gyratonic \emph{pp}-waves deriving a completeness result for these
spacetimes.

\vskip5cm

\section{The metric}
\label{metricforms}

In our analysis we will concentrate on four-dimensional \emph{pp}-wave spacetimes (\ref{pp_metric}), assuming that the flat transverse 2-space spanned by the spatial coordinates ${x^2, x^3}$ is topologically a plane. The gyratonic sources are considered to be localized along (a part of) the axis ${x^2=0=x^3}$, or in a small cylindrical region around this axis. It is thus convenient to introduce polar coordinates by the usual transformation
\begin{equation}\label{transfpol}
x^2=\rho\,\cos\varphi \,,\quad
x^3=\rho\,\sin\varphi \,,
\end{equation}
where ${\rho \in [0,\infty)}$ and the angular coordinate $\varphi$ takes the full range ${\varphi \in [0,2\pi)}$ eliminating ``cosmic strings'' and similar defects along the axis ${\rho=0}$. With the identification\footnote{It eliminates the component $\d u\,\d\rho$. In fact, this is the most reasonable choice to represent the physically relevant quantities in the metric functions, cf. section~\ref{interpretation}.}
\begin{equation}\label{transf}
a_2=-\frac{J}{\rho}\,\sin\varphi \,,\quad
a_3=\frac{J}{\rho}\,\cos\varphi \,,
\end{equation}
implying ${g_{u\rho}=0}$ and ${g_{u\varphi}=J}$, the metric (\ref{pp_metric}) takes the form
\begin{equation}\label{general_metric}
\d s^2 = \d\rho^2+\rho^2\,\d\varphi^2-2\,\d u\,\d r +2\,J(u,\rho,\varphi)\,\d u\,\d\varphi +H(u,\rho,\varphi)\,\d u^2 \,.
\end{equation}
Of course, for consistency, both the metric functions $J$ and $H$ must be $2\pi$-periodic in $\varphi$. In particular, if the functions $J$ and $H$ only depend on the transverse radial coordinate $\rho$ and the retarded time ${u \in (-\infty,\infty)}$, the spacetimes are axially symmetric.

The nonzero Christoffel symbols for the metric (\ref{general_metric}) are
\begin{eqnarray}\label{Christ1}
&&
\Gamma^r_{uu}=-\frac{1}{2}H_{,u}+\frac{1}{2\rho^2}\,J(2J_{,u}-H_{,\varphi}) \,,\qquad
\Gamma^r_{u\rho}=-\frac{1}{2}H_{,\rho}+\frac{1}{2\rho^2}J J_{,\rho} \,,\nonumber\\
&&
\Gamma^r_{u\varphi}=-\frac{1}{2}H_{,\varphi} \,,\qquad
\Gamma^r_{\rho\varphi}=\frac{1}{2\rho}(2J-\rho J_{,\rho}) \,,\qquad
\Gamma^r_{\varphi\varphi}=-J_{,\varphi} \,,\nonumber\\
&&
\Gamma^\rho_{uu}=-\frac{1}{2}H_{,\rho} \,,\qquad
\Gamma^\rho_{u\varphi}=-\frac{1}{2} J_{,\rho} \,,\qquad
\Gamma^\rho_{\varphi\varphi}=-\rho \,,\nonumber\\
&&
\Gamma^\varphi_{uu}=\frac{1}{2\rho^2}(2J_{,u}-H_{,\varphi}) \,,\qquad
\Gamma^\varphi_{u\rho}=\frac{1}{2\rho^2} J_{,\rho} \,,\qquad
\Gamma^\varphi_{\rho\varphi}=\frac{1}{\rho} \,,
\end{eqnarray}
the nontrivial Riemann curvature components are
\begin{eqnarray}\label{Riem1}
&&
R_{u\rho\rho\varphi}=\frac{1}{2\rho}(\rho J_{,\rho\rho}-J_{,\rho}) \,,\qquad
R_{u\varphi\rho\varphi}=\frac{1}{2} J_{,\rho\varphi} \,,\nonumber\\
&&
R_{u\rho u\rho}=-\frac{1}{2}H_{,\rho\rho}+\frac{1}{4\rho^2}(J_{,\rho})^2 \,,
\nonumber\\
&&
R_{u\rho u\varphi}=-\frac{1}{2\rho}(\rho H_{,\rho\varphi}-H_{,\varphi}-\rho J_{,u\rho}+2J_{,u}) \,,
\nonumber\\
&&
R_{u\varphi u\varphi}=-\frac{1}{2}(H_{,\varphi\varphi}+\rho H_{,\rho})+J_{,u\varphi}+\frac{1}{4}(J_{,\rho})^2 \,,
\end{eqnarray}
and the Ricci tensor components read
\begin{eqnarray}\label{Ricci1}
&&
R_{uu}=-\frac{1}{2}\, \triangle H + 2\omega^2 +\frac{1}{\rho^2}\, J_{,u\varphi}\,,\nonumber\\
&&
R_{u\rho}=\frac{1}{\rho} \,\omega_{,\varphi}\,,\qquad
R_{u\varphi}=-\rho \,\omega_{,\rho} \,,
\end{eqnarray}
where
\begin{equation}
\triangle H \equiv H_{,\rho\rho}+\frac{1}{\rho}\, H_{,\rho}+\frac{1}{\rho^2} H_{,\varphi\varphi} \label{Laplace}
\end{equation}
is the 2D flat Laplace operator and, for convenience, the function $\omega$ was defined as
\begin{equation} \label{defomegaJ}
\omega(u,\rho,\varphi) \equiv \frac{J_{,\rho}}{2\rho}\,.
\end{equation}

In Cartesian coordinates, i.e., for the metric form (\ref{pp_metric}), these quantities are given by
\begin{eqnarray}\label{Christ2}
&&
\Gamma^r_{uu}=-\textstyle{\frac{1}{2}}H_{,u}+\delta^{ij}\Big(a_{i,u}-\textstyle{\frac{1}{2}}H_{,i}\Big) a_j \,,\qquad
\Gamma^r_{ui}=-\textstyle{\frac{1}{2}}H_{,i}+\textstyle{\frac{1}{2}}\delta^{jk}(a_{j,i}-a_{i,j})\, a_k \,,\nonumber\\
&&
\Gamma^r_{ij}=-\textstyle{\frac{1}{2}}(a_{i,j}+a_{j,i}) \,,\qquad
\Gamma^i_{uu}=\delta^{ij}\Big(a_{j,u}-\textstyle{\frac{1}{2}}H_{,j}\Big) \,,\qquad
\Gamma^i_{uj}=\textstyle{\frac{1}{2}}\delta^{ik}(a_{k,j}-a_{j,k}) \,,
\end{eqnarray}
\begin{eqnarray}\label{Riem2}
&&
R_{uiuj}=-\textstyle{\frac{1}{2}}H_{,ij}+\textstyle{\frac{1}{2}}(a_{i,uj}+a_{j,ui})+\delta^{kl}a_{[i,k]}\, a_{[j,l]} \,,\nonumber\\
&&
R_{ukij}=\textstyle{\frac{1}{2}}(a_{j,i}-a_{i,j})_{,k} \,,
\end{eqnarray}
\begin{eqnarray}\label{Ricci2}
&&
R_{uu}=-\textstyle{\frac{1}{2}}\delta^{ij}H_{,ij}+\delta^{ij}a_{j,ui}+\textstyle{\frac{1}{2}}\delta^{ij}\delta^{kl}(a_{i,k}-a_{k,i})\, a_{j,l} \,,\nonumber\\
&&
R_{ui}=\textstyle{\frac{1}{2}}\delta^{jk}(a_{j,i}-a_{i,j})_{,k} \,,
\end{eqnarray}
respectively.

\section{Integrating the field equations}
\label{integrating}

In this section we integrate Einstein's equations in the \emph{vacuum region outside} the gyratonic matter source, whose energy-momentum tensor we phenomenologically prescribe to be given by the radiation density $T{_{uu}=\varrho}$ and by the terms ${T_{ui}=j_i}$ representing the spinning character of the gyraton (the remaining components of ${T_{\alpha\beta}}$ are zero) \cite{Bonnor:1970b,Griffiths:1972,FrolovFursaev:2005,FrolovIsraelZelnikov:2005,FrolovZelnikov:2005,FrolovZelnikov:2006,FrolovLin:2006,CaldarelliKlemmZorzan:2007,
YoshinoZelnikovFrolov:2007,KadlecovaZelnikovKrtousPodolsky:2009,KadlecovaKrtous:2010,KrtousPodolskyZelnikovKadlecova:2012}.
When ${j_i=0}$, ${T_{\alpha\beta}}$ reduces to the standard energy-momentum tensor ${T_{\alpha\beta}=\varrho\,k_\alpha k_\beta}$ of pure radiation propagating with the speed of light along the principal null direction ${\boldk=\boldpartial_r}$.

First, it follows from (\ref{Ricci1}) that the Ricci scalar $R$ vanishes. The Einstein field equations ${R_{\alpha\beta}-\frac12 R\, g_{\alpha\beta}=8\pi\,T_{\alpha\beta}}$ (with vanishing cosmological constant) can thus be written as
\begin{eqnarray}
&& \omega_{,\varphi} = 8\pi\rho\, j_\rho \,, \qquad \omega_{,\rho} = -\frac{8\pi}{\rho}\, j_\varphi \,,\label{fieldeqqq1}\\
&& \triangle\, H = 4\,\omega^2 +\frac{2}{\rho^2}\,J_{,u\varphi}-16\pi\varrho \,.\label{fieldeqqq2}
\end{eqnarray}
In general, by specifying the gyratonic matter source $j_i$ one can first integrate equations (\ref{fieldeqqq1}) to obtain $\omega$, and hence $J$ using (\ref{defomegaJ}). Subsequently, prescribing also the radiation density $\varrho$ the metric function $H$ is obtained by solving (\ref{fieldeqqq2}).

In the \emph{vacuum region outside the source}, i.e., assuming ${j_i=0=\varrho}$, we employ the following procedure to obtain a large class of physically interesting explicit solutions. First, from (\ref{fieldeqqq1}) we immediately conclude that $\omega$ must be a function of $u$ only, and using~(\ref{defomegaJ}) we thus obtain the general solution for $J$ in the form
\begin{equation}\label{explicit_metric_function_J}
J = \omega(u)\, \rho^2+\chi (u,\varphi)\,,
\end{equation}
where ${\chi (u,\varphi)}$ is any function, $2\pi$-periodic in $\varphi$. It is convenient to write
\begin{equation}\label{explicit_metric_function_H}
H = \omega^2(u)\,\rho^2+2\,\omega(u)\,\chi (u,\varphi)+ H_0(u,\rho,\varphi)\,,
\end{equation}
since the terms involving $\omega(u)$ in $J$ and $H$ correspond to rigid rotation (they can be generated by the gauge (\ref{gauge1}), (\ref{tildeJH1}) for ${f_{,u}=\omega}$, see below). Substituting (\ref{explicit_metric_function_J}), (\ref{explicit_metric_function_H}) into the remaining field equation (\ref{fieldeqqq2}) we obtain
\begin{equation}\label{Poisson}
\triangle\, H_0 = \rho^{-2}\,\Sigma\,, \qquad \hbox{where}\qquad \Sigma(u,\varphi)\equiv 2(\chi_{,u\varphi}-\omega\,\chi_{,\varphi\varphi})\,.
\end{equation}
A general solution of this Poisson equation can be obtained by Green's function method.

The simplest class of solutions occurs when the function $\chi$ is independent of the angular coordinate $\varphi$, i.e., ${\chi=\chi(u)}$. In such a case ${\Sigma=0}$ and the problem is reduced just to obtain a general homogeneous solution $H_0$,
\begin{equation}\label{fieldeqhomogH}
\triangle \, H_{0}= 0 \,.
\end{equation}
In this way, \emph{any} solution $H_0$ of the Laplace equation (\ref{fieldeqhomogH}) in the flat 2-space (and, of course, their superpositions) generates via (\ref{explicit_metric_function_H}) a particular metric function $H$ representing a possible gyratonic source in the family of \emph{pp}-wave spacetimes. A general solution to the Laplace equation (\ref{fieldeqhomogH}) can conveniently be written by introducing an auxiliary complex variable ${\zeta\equiv\rho\, e^{\iu\varphi}= x^2+\iu x^3}$ in the complete transverse plane, so that the equation becomes ${(H_{0})_{,\zeta\bar\zeta}=0}$. Its solution can be expressed in the form ${H_{0}=F(u,\zeta)+\bar F(u,\bar\zeta)}$ where $F$ is an arbitrary function of $u$ and $\zeta$, holomorphic in $\zeta$. The physically most interesting case is given by a combination\footnote{Terms which are constant and linear in $\zeta$ are omitted since these can be removed by a coordinate transformation.}
\begin{equation}\label{function_f}
F(u,\zeta) = \sum_{m=2}^\infty \alpha_m(u)\,\zeta^m -\mu(u)\log\zeta + \sum_{m=1}^\infty \beta_m(u)\,\zeta^{-m}\,,
\end{equation}
which involves many previously studied non-gyratonic \emph{pp}-wave solutions. Namely, the term ${\alpha_2}$ represents well-known plane gravitational waves \cite{BaldwinJeffrey:1926,Stephani:2003,GriffithsPodolsky:2009}, the higher-order polynomial terms ${\alpha_3, \alpha_4,\ldots}$ correspond to non-homogenous \emph{pp}-waves which exhibit chaotic behaviour of geodesics \cite{PodolskyVesely:1998c,PodolskyVesely:1998d,PodolskyVesely:1999,VeselyPodolsky:2000}, the exceptional logarithmic term is the Aichelburg--Sexl-type solution \cite{AichelburgSexl:1971} (possibly extended \cite{Lessner:1986}), while the inverse-power terms ${\beta_1, \beta_2,\ldots}$ stand for \emph{pp}-waves generated by sources with multipole structure moving along the axis \cite{GriffithsPodolsky:1997,PodolskyGriffiths:1998,Podolsky:1998,FrolovIsraelZelnikov:2005, Podolsky:2002b}.

By putting ${\alpha_m(u)=\frac{1}{2}A_m(u)\, e^{-\iu m\,\varphi'_m(u)}}$ and ${\beta_m(u)=\frac{1}{2}B_m(u)\,e^{\iu m\, \varphi_m(u)}}$, where $A_m$, $B_m$, $\varphi'_m$, $\varphi_m$ are real functions of~$u$, the solution of (\ref{fieldeqhomogH}) corresponding to (\ref{function_f}) reads
\begin{eqnarray}\label{function_H0}
&&H_{0}(u,\rho,\varphi) = \sum_{m=2}^\infty A_m(u)\,\rho^{m}\cos[m(\varphi-\varphi'_m(u))] \nonumber\\
&&\hskip25mm -2\mu(u)\log\rho + \sum_{m=1}^\infty B_m(u)\,\rho^{-m}\cos[m(\varphi-\varphi_m(u))] \,.
\end{eqnarray}
The functions ${A_m(u), B_m(u)}$ give the \emph{amplitudes} of the $m$-components, while ${\varphi'_m(u), \varphi_m(u)}$ determine their \emph{phases}. Observe that the $u$-dependence of ${\varphi'_m(u), \varphi_m(u)}$ enables one to prescribe an arbitrary polarization to any component of the field. The component $A_m(u)$ represents a solution growing as $\rho^{m}$, while the component $B_m(u)$ describes a multipole solution of order $m$, with the monopole solution represented by $\mu(u)$ in (\ref{function_H0}), see also \cite{FrolovIsraelZelnikov:2005}. Indeed, it can be shown that the source of the $m^{\rm th}$ mode is proportional to the $m^{\rm th}$ derivative of the Dirac delta function $\delta(\rho)$ with respect to~$\rho$ \cite{GriffithsPodolsky:1997,PodolskyGriffiths:1998}.

\section{Physical interpretation}
\label{interpretation}

As a next step we will analyze the geometrical and physical meaning of the functions $\omega(u)$ and $\chi (u,\varphi)$ in expressions (\ref{explicit_metric_function_J}), (\ref{explicit_metric_function_H}) for the vacuum metric coefficients $J$ and $H$ of~(\ref{general_metric}).

We start by evaluating the components of the Riemann tensor (\ref{Riem1}) for the explicit solutions
(\ref{explicit_metric_function_J}) and (\ref{explicit_metric_function_H}). The only nonvanishing ones are
\begin{eqnarray}\label{Riemgen}
&&
R_{u\rho u\rho}=-\frac{1}{2}(H_0)_{,\rho\rho} \,,
\nonumber\\
&&
R_{u\rho u\varphi}=-\frac{1}{2\rho}\Big(\rho\, (H_0)_{,\rho\varphi}-(H_0)_{,\varphi}+2(\chi_{,u}-\omega \,\chi_{,\varphi}) \Big) \,,
\\
&&
R_{u\varphi u\varphi}=-\frac{1}{2}\Big((H_0)_{,\varphi\varphi}+\rho (H_0)_{,\rho}-\Sigma\Big) \,.\nonumber
\end{eqnarray}
The spacetimes are thus regular everywhere, except possibly at ${\rho=0}$ and the singularities of the specific solution $H_{0}$.

When ${\chi=\chi(u)}$ implying ${\Sigma=0}$, $H_0$ is given by (\ref{fieldeqhomogH}), in particular (\ref{function_H0}). In the case of spinning multipole particles represented by the terms $\mu$ or ${B_m}$, a curvature singularity occurs at ${\rho=0}$ where the sources of the field are located. For the components $A_m$ the curvature singularities occur at infinity (${\rho=\infty}$) which means that these \emph{pp}-wave spacetimes are not asymptotically flat. It is also interesting to observe that the function $\chi_{,u}(u)$ explicitly occurs in (\ref{Riemgen}), causing a curvature singularity on the axis ${\rho=0}$. On the other hand, the function $\omega(u)$ does \emph{not} occur in the spacetime curvature since ${\chi_{,\varphi}=0}$. Moreover, in this vacuum case, $\omega(u)$ can always be removed by a suitable gauge, as we shall see in the next subsection.

\subsection{Gauge freedom}
\label{gaugefreedom}

We now concentrate on the central issue of the possible removing of the
off-diagonal term $J(u,\varphi,\rho)$ in the metric (\ref{general_metric}), or equivalently
of the terms $a_i(u,x^j)$ in (\ref{pp_metric}), see relation
(\ref{transf}).

First, we consider the gauge freedom
\begin{equation}
\varphi=\tilde \varphi + f(u) \,,\label{gauge1}
\end{equation}
resulting in
\begin{equation}
\tilde J = J + f_{,u}\, \rho^2 \,,\qquad
\tilde H = H + 2J f_{,u}+ f_{,u}^2\, \rho^2\,.\label{tildeJH1}
\end{equation}

Using definition (\ref{defomegaJ}) we immediately conclude that the function
$\omega(u)$ is gauged as ${\tilde\omega=\omega+f_{,u}}$. With an appropriate
choice of $f(u)$ we can thus generate \emph{any} function $\omega(u)$ in $J$, or remove
it by choosing ${f_{,u}=-\omega}$. 
Geometrically, these terms represent just the \emph{rigid rotation of the spacetime}, where $\omega(u)$ is
the corresponding \emph{angular velocity} at different values of $u$. Without
loss of generality, by a suitable gauge \emph{we may thus set} ${\,\omega=0\,}$
in $J$ and $H$ to simplify the metric functions (\ref{explicit_metric_function_J}), (\ref{explicit_metric_function_H}) in the
vacuum region to
\begin{eqnarray}
J &\!=\!& \chi (u,\varphi)\,,\label{Jaftergauge1}\\
H &\!=\!& H_0(u,\rho,\varphi)\,.\label{Haftergauge1}
\end{eqnarray}
In such a most natural ``corotating'' choice of the gauge, $J$ becomes
\emph{manifestly independent} of the radial coordinate $\rho$, corresponding to ${\omega=0}$, see (\ref{defomegaJ}).

As the second step, we employ another gauge freedom of the metric
(\ref{pp_metric}), namely
\begin{equation}
r=\tilde r + g(u,x^i) \,,\label{gauge2cart}
\end{equation}
which implies
\begin{equation}
\tilde a_i = a_i-g_{,i} \,,\qquad
\tilde H = H-2g_{,u}\,.\label{tildeJH2}
\end{equation}
\emph{To achieve} ${\tilde a_i=0}$ for ${\,i,j=2,3\,}$, the function $g$ must be a
\emph{potential} of $a_i$, i.e., ${a_i=g_{,i}}$. A~necessary condition for
this is that the \emph{integrability conditions}
\begin{equation} \label{integrabcondit}
2\,\Omega_{ij} \equiv a_{j,i}-a_{i,j} = g_{,ji}-g_{,ij}=0
\end{equation}
are satisfied.
It is very convenient to express these quantities and relations using 
differential forms defined on the transverse 2-space (spanned by ${x^2,x^3}$),
namely 
the 1-form~${\,\bolda}$, and the 2-form~${\Boldomega}$ as
\begin{equation} \label{omega2formdef}
\bolda\equiv a_i\,\boldd x^i \,, \qquad \Boldomega\equiv \frac{1}{2}\,\boldd
\bolda=\frac{1}{2}\,\Omega_{ij}\,\boldd x^i\wedge \boldd x^j \,.
\end{equation}
The integrability conditions (\ref{integrabcondit}) then translate to
\begin{equation} \label{geomintegrabcondit}
2\,\Boldomega = \boldd \bolda = 0\,.
\end{equation}
It is useful to express the 1-form ${\,\bolda}$ in polar coordinates using
(\ref{transfpol}) and (\ref{transf}) where ${J(u,\rho,\varphi)}$ is the metric
function in (\ref{general_metric}). This leads to the simple expression
\begin{equation} \label{geomintegrabcondit2}
\bolda = J\,\boldd\varphi\,, \qquad \hbox{which implies}\quad \Boldomega
=\frac{1}{2}\,J_{,\rho}\,\boldd \rho\wedge \boldd \varphi
=\omega\,\rho\,\boldd \rho\wedge \boldd \varphi\,,
\end{equation}
see (\ref{defomegaJ}), so that the integrability conditions (\ref{integrabcondit}) turn into the simple single equation
\begin{equation}
J_{,\rho}=0\,. \label{J0}
\end{equation}

Moreover, as we have seen above in (\ref{Jaftergauge1}) this condition can be
assumed to hold without loss of generality in the entire vacuum region. So by the Poincar\'e lemma,
the closed form $\bolda$ is locally exact, i.e., locally in the vacuum region there exists a suitable
function $g$ such that ${\bolda=\boldd g}$. In view of (\ref{tildeJH2}), the corresponding gauge transformation (\ref{gauge2cart}) then explicitly
removes all the components $\tilde a_i$ from the metric of the form (\ref{pp_metric}), that is ${\bolda=0}$.

However, since we have assumed the source to be located along (a part of) the axis
the vacuum region is \emph{not contractible} and hence the closed form $\bolda$ is \emph{not globally exact},
which means that even in the vacuum region we cannot globally remove the off-diagonal terms in the
metric. Clearly, the properties of possible gyratons are related to the cohomology of
the vacuum region.

Summing up, when considering \emph{pp}-wave spacetimes with gyratonic sources located along $\rho=0$
it is \emph{not only preferable but in fact necessary to keep the off-diagonal terms in the metric}
and employ them to express the relevant physical parameter (namely the angular momentum of the source) in the most
efficient way which we will do next. To this end we consider the contour integral
\begin{equation} \label{integral}
\oint_C \bolda=\oint_C a_i \,\boldd x^i= \oint_C J\,\boldd \varphi
\,,
\end{equation}
where $C$ is an arbitrary contour in the transverse 2-space running around the
axis ${\rho=0}$ (once and counterclockwise). In fact, by (\ref{geomintegrabcondit}) it is
independent of the choice of the contour~$C$ in the vacuum region, and it is also gauge-independent with respect to (\ref{tildeJH2}).

\subsection{Energy and angular momentum of the source}
\label{energyANDmomentum}

Now, following previous works of Bonnor and Frolov with collaborators \cite{Bonnor:1970b, FrolovFursaev:2005}, we will relate the metric functions $H$ and $J$ to the principal physical properties of the source, namely its total energy and angular momentum. In particular, we will prove that the integral (\ref{integral}) directly determines the angular momentum density of the gyratonic source.

\begin{figure}[t]
\begin{center}
\includegraphics[scale=0.66]{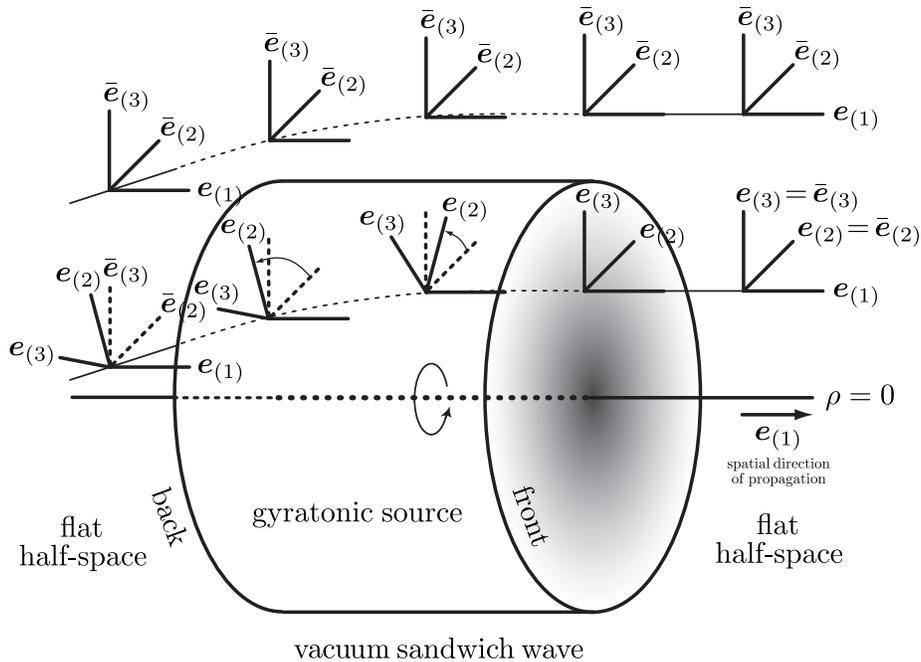}%
\caption{Schematic picture of the localized gyratonic source generating a vacuum sandwich gravitational wave. It also indicates
the dragging effect which causes the parallelly propagated interpretation Cartesian frames ${\{\bolde_{(1)}, \bolde_{(2)}, \bolde_{(3)} \}}$ to rotate inside the gyraton with the angular velocity ${\dot u\,\omega(u,\rho,\varphi)}$ with respect to the background frames ${\{\bar{\bolde}_{(1)}, \bar{\bolde}_{(2)}, \bar{\bolde}_{(3)} \}}$. In the region outside the source it is always possible to set ${\omega=0}$.}
\label{fig1}
\end{center}
\end{figure}

In the linearized theory when the gravitational field is weak the \emph{total mass-energy}~$M$ and \emph{total angular momentum}~$J^{\mu\nu}$ (relative to the origin of coordinates) on the spacelike hypersurface of constant time $t$ are given by \cite{MisnerThorneWheeler:1973}
\begin{equation} \label{mass angmom_density_definition}
M \equiv \int\!\!\!\int\!\!\!\int T^{tt}\, \d x \d y \d z \,,
\qquad
J^{\mu\nu} \equiv \int\!\!\!\int\!\!\!\int (x^\mu T^{\nu t}- x^\nu T^{\mu t})\, \d x \d y \d z \,,
\end{equation}
respectively, where ${x^\mu=(t,x,y,z)}$ are Minkowski background coordinates. In this section it is assumed that the energy-momentum tensor $T_{\mu\nu}$ of the source is localized in a cylindrical region of radius $R$ around the axis ${\rho=0}$, with a finite length~$L$ in the $z$-direction, as shown in figure~\ref{fig1}. As is standard, we assume that the cylindrical source region is matched to the external vacuum region in a $C^1$-way.

For negligible metric functions $H$ and $a_i$ the metric (\ref{pp_metric}) approaches flat Minkowski space with the coordinates ${u=\ssqrt(t-z)}$, ${r=\ssqrt(t+z)}$, ${x^2=x}$, ${x^3=y}$. The nontrivial gyratonic components are $T{_{uu}=\varrho}$ and ${T_{ui}=j_i}$, yielding $T{^{tt}=\frac{1}{2}\varrho}$, ${T^{xt}=-\ssqrt\, j_x}$ and ${T^{yt}=-\ssqrt\, j_y}$. Since $t$ is fixed we can substitute ${\d z}$ by ${-\sqrt2\,\d u}$ in which, however, the negative sign
is effectively compensated by the fact that the boundary value ${z=0}$ corresponds to ${u=\ssqrt t}$ while ${z=L}$ corresponds to ${u=\ssqrt(t-L)<\ssqrt t}$. Therefore,
\begin{equation} \label{mass angmom_density_evaluation}
M = \ssqrt\int\!\!\!\int\!\!\!\int \varrho\, \d x \d y \d u \,,
\qquad
J^{xy} = \int\!\!\!\int\!\!\!\int (y j_x-x j_y)\, \d x \d y \d u \,.
\end{equation}
The function
\begin{equation} \label{massdensity}
{\cal M}(u) = \ssqrt\int\!\!\!\int_S \varrho\, \d x \d y \,,
\end{equation}
where the surface integral is taken over the disc $S$ of radius $R$ in the transverse space,
thus represents the \emph{mass-energy density} of the source as a function of the retarded time~$u$ such that ${M = \int\! {\cal M}(u)\, \d u}$, while the \emph{angular momentum density} of the gyratonic source is given by
\begin{equation} \label{angmomdensity}
{\cal J}(u) = \int\!\!\!\int_S (y j_x - x j_y)\, \d x \d y \,,
\end{equation}
so that ${J^{xy} = \int\! {\cal J}(u)\, \d u\,}$.

It is now seen from the field equation (\ref{fieldeqqq2}) that the mass-energy density ${\cal M}(u)$ of the source is determined by the metric function $H$, namely by the surface integral of ${\triangle\, H}$ in the transverse space (in the gauge ${\omega=0}$ and for ${J_{,\varphi}=0}$).

Interestingly, the angular momentum density ${\cal J}(u)$ is directly determined by the contour integral of the metric function $J$ or, equivalently, $a_i$. Indeed, expressing (\ref{angmomdensity}) in polar coordinates and employing the second field equation (\ref{fieldeqqq1}) it becomes
\begin{equation} \label{angmomdensitypol}
{\cal J}(u) = -\int\!\!\!\int_S j_\varphi\, \rho\,\d \rho\,\d \varphi = \frac{1}{8\pi}\int_0^{2\pi}\!\!\Big(\int_0^R \omega_{,\rho}\, \rho^2\d \rho\Big)\,\d \varphi \,.
\end{equation}
Using integration by parts and assuming that ${\omega\rho^2}$ vanishes at ${\rho=0}$, the inner integral can be rewritten as
\begin{equation} \label{inneritn}
\omega(R)\, R^2- 2\int_0^R \omega\, \rho\,\d \rho\,.
\end{equation}
However, ${\omega(R)=0}$ because on the cylindrical boundary ${\rho=R}$ of the gyratonic source the metric function $J$ and its derivative ${J_{,\rho}=2\omega\rho}$, cf.~(\ref{defomegaJ}), are continuously joined to the external vacuum region in which ${\omega=0}$ everywhere (using the gauge freedom removing the rigid rotation). We thus obtain
\begin{equation} \label{angmomdensityfin1}
{\cal J}(u) = -\frac{1}{4\pi}\int\!\!\!\int_S \omega\, \rho\,\d \rho\,\d \varphi \,,
\end{equation}
which can be reexpressed in a geometric way using expression (\ref{geomintegrabcondit2}) for the 2-form $\Boldomega$ as
\begin{equation} \label{integralStokes}
{\cal J}(u) =
-\frac{1}{4\pi}\int\!\!\!\int_S \omega\, \rho\,\boldd \rho\wedge \boldd \varphi=
-\frac{1}{4\pi}\int\!\!\!\int_S \Boldomega \,.
\end{equation}
Now, employing (\ref{omega2formdef}) and the Stokes theorem we obtain
\begin{equation} \label{integralStokesfinal}
{\cal J}(u) =
-\frac{1}{8\pi}\int\!\!\!\int_S \boldd \bolda = -\frac{1}{8\pi}\oint_C \bolda\,,
\end{equation}
where ${C=\partial S}$ is the outer contour. Therefore, the gauge-independent contour integral (\ref{integral}) directly determines the angular momentum density of the gyratonic source.

Moreover, from (\ref{integralStokes}) we conclude that if ${\Boldomega=0}$ everywhere in the whole spacetime, the integrals (\ref{integral}) vanish for \emph{any} closed contour $C$ and there is no gyraton. The presence of the gyraton is identified by ${\Boldomega\not=0}$ in some region, for example along (a part of) the axis. This necessarily implies ${{\cal J}(u)\not=0}$, so that the angular momentum is nonvanishing.

In particular, for the simplest gyraton \cite{Bonnor:1970b,Griffiths:1972,FrolovIsraelZelnikov:2005,FrolovFursaev:2005,YoshinoZelnikovFrolov:2007} located at ${\rho=0}$ given by ${J=J(u)}$ only, we have ${\bolda = J(u)\,\boldd\varphi}$ in the external vacuum region. This closed form $\bolda$ is not globally exact, hence it can not be \emph{globally} removed and ${{\cal J}(u)=-\frac{1}{4} J(u)}\not=0$ for any loop $C$ around ${\rho=0}$ as in (\ref{integral}). Hence, the source is gyrating.

\subsection{Interpretation frame}
\label{interpretationframe}

To support this conclusion---and to enable a further analysis---it is convenient to introduce a suitable \emph{interpretation orthonormal frame} ${\{\bolde_a(\tau) \}}$. At any point along an arbitrary (future-oriented) timelike geodesic ${\gamma(\tau)}$, where $\tau$ is the proper time, this defines an observer's framework in which physical measurements are made and interpreted. The timelike vector is identified with the \emph{velocity vector} of the observer, ${\bolde_{(0)}=\boldu}$, while ${\bolde_{(1)},\bolde_{(2)},\bolde_{(3)}}$ are perpendicular \emph{spacelike unit vectors} which form its local Cartesian basis in the hypersurface orthogonal to $\boldu$,
${\bolde_a \cdot \bolde_b \equiv g_{\alpha\beta}\, e_a^{\>\alpha} e_b^{\>\beta}=\hbox{diag}(-1,1,1,1)}$.
It is also convenient to introduce an associated \emph{null frame} ${\{\boldk, \boldl, \bolde_{(2)}, \bolde_{(3)} \}}$ by the relations
\begin{equation}\label{NullFrame}
\boldk=\ssqrt(\boldu+\bolde_{(1)})\,, \quad \boldl=\ssqrt(\boldu-\bolde_{(1)}) \,.
\end{equation}
Thus $\boldk$ and $\boldl$ are future-oriented null vectors, while $\bolde_{(i)}$ for ${\,i=2,3\,}$ are spatial unit vectors orthogonal to them:
${\boldk\cdot\boldl=-1}$, ${\bolde_{(i)}\cdot\bolde_{(j)}=\delta_{ij}}$, ${\boldk\cdot\boldk=0=\boldl\cdot\boldl}$, ${\boldk\cdot\bolde_{(i)}=0=\boldl\cdot\bolde_{(i)}}$. In view of definition (\ref{NullFrame}), the spatial vector ${\bolde_{(1)}=\sqrt{2}\,\boldk-\boldu}$ is privileged, and we will refer to it as the \emph{longitudinal vector}. The vectors ${\bolde_{(2)},\bolde_{(3)}}$ will be called the \emph{transverse vectors}.

For the \emph{pp}-wave metric (\ref{pp_metric}) such an \emph{interpretation null frame}, adapted to any geodesic observer with the four-velocity ${\boldu=\dot{r}\,\boldpartial_r+\dot{u}\,\boldpartial_u+\dot{x}^2\,\boldpartial_{x^2} + \dot{x}^3\,\boldpartial_{x^3}}$, reads
\begin{eqnarray}\label{null_frame_pp}
\boldk &\!=\!& \frac{1}{\sqrt{2}\,\dot{u}}\,\boldpartial_r \,, \nonumber \\
\boldl &\!=\!& \bigg(\sqrt{2}\,\dot{r}-\frac{1}{\sqrt{2}\,\dot{u}}\,\bigg)\,\boldpartial_r + \sqrt{2}\,\dot{u}\,\boldpartial_u + \sqrt{2}\,\dot{x}^2\,\boldpartial_{x^2} + \sqrt{2}\,\dot{x}^3\,\boldpartial_{x^3} \,, \nonumber\\
\bolde_{(i)} &\!=\!& \bigg(a_k+\frac{\dot{x}^j}{\dot{u}}\delta_{jk}\bigg)e_{(i)}^{\>k}\,\boldpartial_r + e_{(i)}^{\>2}\,\boldpartial_{x^2} + e_{(i)}^{\>3}\,\boldpartial_{x^3} \,,
\end{eqnarray}
where ${\,\delta_{kl}\,e_{(i)}^{\>k}\,e_{(j)}^{\>l}=\delta_{ij}\,}$, and the dot denotes differentiation with respect to~$\tau$. In view of (\ref{Christ1}) or (\ref{Christ2}) it immediately follows that ${\dot u=\,}$const.\ along any geodesic.
Notice that $\boldk$ is proportional to the privileged null vector field $\boldpartial_r$ which is covariantly constant. Its spatial projection is oriented along the longitudinal vector ${\bolde_{(1)}}$, which thus represents the \emph{propagation direction} of gravitational waves and the gyratonic source, whereas ${\bolde_{(2)}}$, ${\bolde_{(3)}}$ span the transverse 2-space at any $\tau$.

Let us investigate the behaviour of the interpretation frame under the gauge (\ref{gauge1}), ${\varphi=\tilde\varphi+f(u)}$ with ${f_{,u}=-\omega}$, followed by the gauge (\ref{gauge2cart}). The latter, expressed in polar coordinates with ${\rho=\tilde\rho}$, takes the form ${r=\tilde r + g(u,\tilde\varphi)}$, implying ${\tilde J = J-g_{,\tilde\varphi} \,}$ and ${\tilde H = H-2g_{,u} \,}$. Using the relations (\ref{transfpol}), (\ref{transf}), and taking the most convenient choice
\begin{equation}\label{cylindrchoice}
e_{(2)}^{\>2}=\cos\varphi\,,\quad e_{(2)}^{\>3}=\sin\varphi\,,\qquad\qquad e_{(3)}^{\>2}=-\sin\varphi\,,\quad e_{(3)}^{\>3}=\cos\varphi
\end{equation}
in (\ref{null_frame_pp}), we obtain
\begin{eqnarray}
\boldk &\!=\!& \frac{1}{\sqrt{2}\,\dot{u}}\,\boldpartial_{\tilde r} \,, \label{null_frame_pp_gaugedk}\\
\boldl &\!=\!& \bigg(\sqrt{2}\,{\dot{\tilde r}}-\frac{1}{\sqrt{2}\,\dot{u}}\,\bigg)\,\boldpartial_{\tilde r} + \sqrt{2}\,\dot{u}\,\boldpartial_u + \sqrt{2}\,\dot{\rho}\,\boldpartial_{\rho} + \sqrt{2}\,\dot{\tilde\varphi}\,\boldpartial_{\tilde\varphi} \,, \label{null_frame_pp_gaugedl}\\
\bolde_{(2)} &\!=\!& \frac{\dot{\rho}}{\dot{u}}\,\boldpartial_{\tilde r} + \boldpartial_{\rho} \,,\label{null_frame_pp_gauged2}\\
\bolde_{(3)} &\!=\!& \bigg(\,\frac{\tilde J}{\rho}+\rho\,\frac{\dot{{\tilde\varphi}}}{\dot{u}}\,\bigg)\,\boldpartial_{\tilde r} + \frac{1}{\rho}\,\boldpartial_{\tilde\varphi} \,,\label{null_frame_pp_gauged3}
\end{eqnarray}
where
\begin{equation}\label{Jphys}
\tilde J = J-\omega\, \rho^2-g_{,\tilde\varphi} \,.
\end{equation}
This gives the interpretation frame adapted to the polar coordinates of the metric (\ref{general_metric}), for which ${\,\boldpartial_{\rho}\,}$ and ${\,\frac{1}{\rho}\,\boldpartial_{\tilde\varphi}\,}$ are the natural radial and axial unit vectors in the transverse space.
The form of the interpretation frame is clearly gauge invariant, with the \emph{physical part} of the off-diagonal metric function determined by (\ref{Jphys}). In fact, we can always \emph{fix the most suitable gauge} of coordinates (and thus the ``canonical'' frame) in such a way that $\tilde J$ does \emph{not} contain the rigid rotation and the trivial part generated by the potential $g$.

For \emph{static} geodesic observers with ${\dot{\rho}=0=\dot{\tilde\varphi}}$, the expressions (\ref{null_frame_pp_gauged2}), (\ref{null_frame_pp_gauged3}) simplify to
\begin{equation}\label{staticframe}
\,\bolde_{(2)} = \boldpartial_{\rho}\,,\qquad
\bolde_{(3)} = \frac{1}{\rho}\,(\tilde J\,\boldpartial_{\tilde r}+\boldpartial_{\tilde\varphi})\,.
\end{equation}
Interestingly, the single function $\tilde J$ directly enters (only) the expressions (\ref{null_frame_pp_gauged3}) or (\ref{staticframe}) for the \emph{axial} vector $\bolde_{(3)}$, distinguishing thus the usual \emph{pp}-wave case ${\tilde J=0}$ from the case ${\tilde J\not=0}$ that involves the spinning gyraton source. Indeed, for the simplest gyraton \cite{Bonnor:1970b,FrolovIsraelZelnikov:2005,FrolovFursaev:2005} given by ${\tilde J=\chi(u)\not=0}$, the invariant contour integral (\ref{integral}) around ${\rho=0}$ gives the angular momentum density ${{\cal J}(u)=-\frac{1}{4}\,\chi(u)}$, see the end of subsection~\ref{energyANDmomentum}.

\subsection{Analysis of the dragging effect}
\label{rotation}

Now we will demonstrate that the off-diagonal metric functions $a_i$ in (\ref{pp_metric}), or equivalently the function $J$ in (\ref{general_metric}), directly encode the rotational ``dragging'' effect of the spinning source on the spacetime. To this end we employ the canonical orthonormal frame ${\{\bolde_a\}}$ adapted to a (timelike) geodesic observer and the associated null frame ${\{\boldk, \boldl, \bolde_{(2)}, \bolde_{(3)} \}}$. Using their mutual relation (\ref{NullFrame}), we can prove that the \emph{interpretation null frame} (\ref{null_frame_pp}) \emph{is parallelly transported} along a timelike geodesic $\gamma(\tau)$ in the spacetime (\ref{pp_metric}) \emph{if, and only if,}
\begin{equation}
\frac{\d e_{(i)}^{\>k}}{\d\tau} = {\Omega^{\,k}}_{l} \, e_{(i)}^{\>l}\,, \label{condit1}
\end{equation}
${i,k,l=2,3\,}$, in which ${{\Omega^{\,k}}_{l}}$ are elements of the antisymmetric ${2\times2}$ matrix
\begin{equation}
{\Omega^{\,k}}_{l}(\tau) \equiv \dot{u}\,\delta^{kj}\,\Omega_{jl} \,, \label{omegadef}
\end{equation}
where ${\,\Omega_{jl} = \frac{1}{2}(a_{l,j}-a_{j,l})}$ are the components of the 2-form ${\Boldomega=\frac{1}{2}\boldd \bolda}$, see (\ref{integrabcondit}), (\ref{omega2formdef}). The only nontrivial matrix element is ${{\Omega}^{\,2}_{\ 3}=-{\Omega}^{\,3}_{\ 2}\,}$, so that ${{\Omega}^{\,2}_{\ 3}(\tau)={\dot u}\,\Omega_{23}=\frac{1}{2}\,\dot{u}\,(a_{3,2}-a_{2,3})}$,
evaluated along $\gamma(\tau)$, where ${\dot u=\,}$const. Moreover, using (\ref{transfpol}), (\ref{transf}), (\ref{defomegaJ}) we see that
\begin{equation} \label{omegaJ2}
\Omega_{23} = \frac{J_{,\rho}}{2\rho} = \omega(u,\rho,\varphi)\,.
\end{equation}

To prove (\ref{condit1}), we use the fact that the vector $\boldk$ is covariantly constant and thus parallely transported, and therefore ${\boldl=\sqrt{2}\,\boldu-\boldk}$ is also parallelly transported. It only remains to ensure that $\frac{\D e_{(i)}^{\>\mu} }{\d\tau} =0$ for ${\,i=2,3\,}$. By using (\ref{null_frame_pp}), (\ref{Christ2}) we obtain ${\frac{\D e_{(i)}^{\>u} }{\d\tau} =0}$. The spatial components yield the condition (\ref{condit1}), (\ref{omegadef}). Finally, from the derivative of the condition ${\boldu\cdot\bolde_{(i)}=0\,}$ it follows that ${\,g_{\alpha\beta}\, u^\alpha \frac{\D e_{(i)}^{\>\beta} }{\d\tau}=0}$. Since ${\frac{\D e_{(i)}^{\>u} }{\d\tau} =0=\frac{\D e_{(i)}^{\>k} }{\d\tau} }$ we obtain ${\,g_{\alpha r}\, u^\alpha \frac{\D e_{(i)}^{\>r} }{\d\tau}=-\dot u \frac{\D e_{(i)}^{\>r} }{\d\tau} =0\Rightarrow\frac{\D e_{(i)}^{\>r} }{\d\tau} =0}$, which completes the argument (recalling that when ${\dot u=0}$ the geodesic cannot be timelike).

If ${\Boldomega=0}$ \emph{everywhere} in the spacetime, the metric functions~$a_i$ can be globally removed by (\ref{gauge2cart}) and (\ref{tildeJH2}), hence there is \emph{no gyraton}. From (\ref{condit1}), (\ref{omegadef}) it then follows that the coefficients ${e_{(i)}^{\>k}}$ of the parallelly propagated interpretation frame are \emph{just constants}. It is thus natural to consider a \emph{reference Cartesian basis} ${\bar{\bolde}_{(2)}}$, ${\bar{\bolde}_{(3)}}$ given by the simplest choice ${{\bar e}_{(i)}^{\>k}=\delta_i^k}$. The corresponding null reference frame ${\{\boldk, \boldl, \bar{\bolde}_{(2)}, \bar{\bolde}_{(3)} \}}$, where
\begin{eqnarray}\label{null_ref_ frame_pp}
\bar{\bolde}_{(i)} &\!\equiv\!&
\bigg(a_i+\delta_{ij}\frac{\dot{x}^j}{\dot{u}}\bigg)\,\boldpartial_r + \boldpartial_{x^i} \,,
\end{eqnarray}
is parallelly propagated along all timelike geodesics in any non-gyratonic \emph{pp}-wave spacetime and, in particular, in Minkowski background (with the usual choice ${a_i=0}$).

It is possible (and useful) to introduce the reference frame (\ref{null_ref_ frame_pp}) along any geodesic in the \emph{general} \emph{pp}-wave spacetime with a gyraton encoded by \emph{nontrivial} functions~$a_i$. Of course, it remains parallelly propagated in the flat Minkowski regions in front and behind the sandwich/impulsive wave. Interestingly, this is also true in the \emph{vacuum region outside the gyratonic source} (after removing the global rigid rotation function $\omega(u)$ by a suitable gauge (\ref{gauge1}), cf. the canonical choice allowed by (\ref{Jphys})), as indicated in figure~\ref{fig1}. On the other hand, \emph{inside the gyratonic source} the metric components $a_i$ are such that ${\Boldomega\not=0}$. The reference frame (\ref{null_ref_ frame_pp}) thus does \emph{not} propagate parallelly in the source region of the spacetime because the right-hand side of (\ref{condit1}) is nonzero. Instead, it is the frame ${\{\boldk, \boldl, \bolde_{(2)}, \bolde_{(3)} \}}$ given by (\ref{null_frame_pp}) that is parallelly transported, provided (\ref{condit1}) is satisfied.

Because both the bases ${{\bar{\bolde}_{(2)}},\bar{\bolde}_{(3)}}$ and ${\bolde_{(2)},\bolde_{(3)}}$ in the transverse 2-space are normalized to be perpendicular unit vectors, they must be related by a linear transformation
\begin{equation}\label{orthotransf}
\bolde_{(i)}= {A^{\,j}}_i \,\bar{\bolde}_{(j)} \,,
\end{equation}
where ${\,{A^{\,j}}_i(\tau)\,}$ are elements of an \emph{orthonormal} ${2\times2}$ matrix.
The antisymmetric matrix ${{{\Omega}^{\,k}}_{l}(\tau)}$
introduced in (\ref{omegadef}) is thus the \emph{angular velocity of rotation} of the parallelly transported interpretation basis $\bolde_{(i)}$, given by (\ref{null_frame_pp}), with respect to the reference basis $\bar{\bolde}_{(k)}$, introduced in (\ref{null_ref_ frame_pp}). Indeed, for ${{\bar e}_{(j)}^{\>k}=\delta_j^k}$ it follows from (\ref{orthotransf}) that ${e_{(i)}^{\>k}={A^{\,k}}_i}$.
Differentiating this relation with respect to $\tau$ and using (\ref{condit1}) we get ${\frac{\d}{\d\tau}{A^{\,k}}_i={\Omega^{\,k}}_{j} \,{A^{\,j}}_i}$. Multiplication by the inverse matrix ${{(A^{-1})^{\,i}}_l={(A^{T})^{\,i}}_l}$ yields
\begin{equation}\label{angulvelocitymatrix}
{\Omega}^{\,k}_{\ l} = {\big(\,{\textstyle\frac{\d}{\d\tau}}A\,\big)^{\,k}}_i\,{\big(\,A^{T}\,\big)^{\,i}}_l\,.
\end{equation}
It is well known \cite{AbrahamMarsden:1978,JoseSaletan:1998} that this is the antisymmetric angular velocity matrix corresponding to the rotation described by $A$.

This enables us to physically interpret the metric component $J$ in the \emph{pp}-wave metric (\ref{general_metric}): \emph{inside the gyratonic source it causes the ``rotation dragging effect''} on parallelly propagated frames, as shown in figure~\ref{fig1}. Specifically, the parallelly transported frame (\ref{null_frame_pp}) rotates in the 2-space spanned by ${\bolde_{(2)},\bolde_{(3)}}$ with the angular velocity of rotation
${ {\Omega}^{\,2}_{\ 3}(\tau)={\dot u}\, \omega\Big(u(\tau),\rho(\tau),\varphi(\tau)\Big)}$,
where ${\omega = J_{,\rho}/2\rho\,}$. Such rotation is measured with respect to the background reference frame ${\{\boldk, \boldl, \bar{\bolde}_{(2)}, \bar{\bolde}_{(3)} \}}$, where ${\boldk, \boldl}$ are the same as in (\ref{null_frame_pp}) while ${\bar{\bolde}_{(2)}, \bar{\bolde}_{(3)}}$ are defined in (\ref{null_ref_ frame_pp}), which is the most natural choice when ${J=0}$.

Since we are primarily interested in impulsive or sandwich \emph{pp}-waves which have compact supports (finite duration), the frames ${\bar{\bolde}_{(i)}}$ are well defined both ``in front'' and ``behind'' the wave, see figure~\ref{fig1}. Moreover, when the gyratonic source is localized in a cylindrical region of radius $R$ around ${\rho=0}$, the vacuum region \emph{outside} such a source \emph{globally} admits the parallelly transported frame ${\{\boldk, \boldl, \bar{\bolde}_{(2)}, \bar{\bolde}_{(3)} \}}$. It forms the reference frame of distant observers, with respect to which the parallelly transported frame (\ref{null_frame_pp}) \emph{inside} the source rotates.

Within the gyratonic source it follows from (\ref{fieldeqqq1}) that
\begin{equation} \label{omegaingyraton}
\omega(u,\rho,\varphi)=-8\pi\int\frac{j_\varphi}{\rho}\,\d\rho\,.
\end{equation}
At any fixed $u$ and $\varphi$, the function $\omega$ is depending on $\rho$, so that the angular velocity ${{\dot u}\, \omega}$ depends on the radial distance from the axis. It is thus a \emph{differential rotation} which (in contrast to the rigid one) cannot be removed by a gauge. For the \emph{Bonnor solution} \cite{Bonnor:1970b} there is ${\omega\propto (R-\rho)}$ inside the cylindrical gyratonic source, i.e., $\omega$ linearly decreases to zero at its outer boundary ${\rho=R}$, and ${\omega=0}$ everywhere in the external vacuum region. This is fully consistent with the integrability conditions (\ref{integrabcondit}) for the 2-form ${\Boldomega=\omega\,\rho\,\boldd \rho\wedge \boldd \varphi}$, see (\ref{geomintegrabcondit2}). Such conditions are obviously valid in the external vacuum region ${\rho\ge R}$, but inside the gyratonic source ${0\le\rho<R}$ there is ${\Boldomega\not = 0}$, and by (\ref{integralStokes}) the gyratonic source has a nonvanishing angular momentum density given by
${{\cal J}(u)\not=0}$.

\subsection{The field scalars}
\label{NPscalars}

To determine the algebraic structure of the general spacetime (\ref{general_metric}), it is important to evaluate the nontrivial Newman--Penrose scalars which are components of the gravitational and gyratonic matter fields in a suitable null frame. We employ the frame
\begin{eqnarray}
\boldk &\!=\!& \boldpartial_{r} \,, \label{null_frame_speck}\\
\boldl &\!=\!& \boldpartial_u + {\textstyle \frac{1}{2}}H\,\boldpartial_{r} \,, \label{null_frame_specl}\\
\boldm &\!=\!& {\textstyle \frac{1}{\sqrt2}}(\bolde_{(2)}+\iu\bolde_{(3)})=
{\textstyle \frac{1}{\sqrt2}}\Big(\,\boldpartial_{\rho}
+\frac{\iu}{\rho}\,(\boldpartial_{\varphi}+J\,\boldpartial_{r}) \Big) \,,
\label{null_frame_specm}
\end{eqnarray}
which is a particular case of (\ref{null_frame_pp_gaugedk})--(\ref{null_frame_pp_gauged3})
with ${\dot{\rho}=0=\dot{\varphi}}$, ${\dot{u}=\frac{1}{\sqrt2}}$, where we have dropped the tildes. Projecting the Weyl tensor components (using (\ref{Riem1}), (\ref{Ricci1}) and ${R=0}$) we obtain
\begin{eqnarray}
&&
\Psi_4=-\frac{1}{4}\bigg( H_{,\rho\rho}-\frac{1}{\rho}H_{,\rho}-\frac{1}{\rho^2}H_{,\varphi\varphi}\bigg)
-\frac{1}{2\rho^2}\,J_{,u\varphi}
\nonumber\\
&& \hspace{11mm}+\frac{\iu}{2\rho^2}\bigg(\rho H_{,\rho\varphi}-H_{,\varphi}-\rho J_{,u\rho}+2J_{,u} \bigg) \,,\label{Psi4}\\
&&
\Psi_3=-\frac{1}{2\sqrt2}\bigg(\, \frac{1}{\rho}\,\omega_{,\varphi}+\iu \omega_{,\rho}\bigg)
=-\frac{4\pi}{\sqrt2}\bigg(\, j_\rho-\frac{\iu}{\rho}\, j_\varphi\bigg)\,,
\label{Psi3}
\end{eqnarray}
while the nonvanishing Ricci tensor components read
\begin{eqnarray}
&&
\Phi_{22}=-\frac{1}{4}\, \triangle H + \omega^2 +\frac{1}{2\rho^2}\, J_{,u\varphi}
\,,\label{Phi22}\\
&&
\Phi_{12}=-\bar\Psi_3 \,,\label{Phi12}
\end{eqnarray}
generalizing the results presented in section 18.5 of \cite{GriffithsPodolsky:2009}. The spacetime inside the gyratonic source is thus of Petrov type~III. Notice the interesting fact that the \emph{gyrating matter component} $\Phi_{12}$ \emph{is uniquely connected to the gravitational field component} $\Psi_3$. In particular, they vanish simultaneously, so that the spacetime is of type~N if, and only if, there is no gyratonic matter in the given region. In such a case, the only nontrivial Newman--Penrose scalars are $\Psi_4$ and $\Phi_{22}$ (representing pure radiation matter field).

Moreover, in the \emph{vacuum region outside the gyratonic source} the field equations (\ref{fieldeqqq1}), (\ref{fieldeqqq2}) with ${j_i=0=\varrho}$ guarantee that ${\Phi_{12}=0=\Phi_{22}}$ and ${\Psi_3=0}$. The gravitational field is of type~N with the scalar $\Psi_4$, which simplifies using (\ref{explicit_metric_function_J})--(\ref{Poisson}) to
\begin{equation}
\Psi_4=-\frac{1}{2}\,(H_0)_{,\rho\rho}
+\frac{\iu}{2\rho^2}\Big(\,\rho\, (H_0)_{,\rho\varphi}-(H_0)_{,\varphi}+2(\chi_{,u}-\omega \,\chi_{,\varphi}) \Big) \,.\label{Psi4vac}
\end{equation}
It can be observed that the real part of this curvature component is given by the second derivative of the function $H_0$ in the radial direction $\rho$, while the imaginary part is determined by its derivative in the angular direction $\varphi$ and the specific derivatives of $\chi$.

\section{Geodesic deviation}
\label{frame}

Further physical properties of the gyratonic \emph{pp}-wave spacetimes can be obtained by studying the specific deviation of nearby (timelike) geodesics. Such relative motion of free test particles is described by the equation of geodesic deviation \cite{MisnerThorneWheeler:1973}. To obtain invariant results, we employ the natural orthonormal frame introduced in subsection~\ref{interpretationframe}. As summarized, e.g., in \cite{PodolskySvarc:2012}, the geodesic deviation equation then takes the form
\begin{equation}\label{InvGeoDev}
\ddot Z^{(\rm{i})}= R^{(\rm{i})}_{\ \> (0)(0)(\rm{j})}\,Z^{(\rm{j})} \,,
\end{equation}
where ${Z^{(\rm{i})}(\tau)\equiv\bolde^{(\rm{i})}\cdot\boldZ\,}$ for ${\,\rm{i},\rm{j}=1,2,3\,}$ are spatial (Cartesian) frame components of the separation vector~$\boldZ(\tau)$ determining the \emph{relative spatial position} of two test particles, the \emph{physical relative acceleration} is given by ${\ddot Z^{(\rm{i})}(\tau) \equiv \bolde^{(\rm{i})}\cdot \Big(\frac{\D^2}{\d\tau^2}\boldZ\Big)}$, and the relevant frame components of the Riemann tensor are ${R^{(\rm{i})}_{\ \> (0)(0)(\rm{j})}=R_{(\rm{i})(0)(0)(\rm{j})}\equiv R_{\mu\alpha\beta\nu} \,e^\mu_{(\rm{i})}u^\alpha u^\beta e^\nu_{(\rm{j})}}$.

In view of (\ref{Riemgen}) with (\ref{fieldeqhomogH}), the only nontrivial components ${R_{(\rm{i})(0)(0)(\rm{j})}}$ in the reference frame (\ref{null_frame_pp_gaugedk})--(\ref{null_frame_pp_gauged3}) where ${\bolde_{(0)}=\boldu=\ssqrt(\boldk+\boldl)}$, simplified by the gauge (\ref{gauge1}), (\ref{gauge2cart}), are
${ R_{(2)(0)(0)(2)} = -R_{(3)(0)(0)(3)} \equiv {\cal A}_+}$,
${ R_{(2)(0)(0)(3)} \equiv {\cal A}_\times}$, in which
\begin{eqnarray}\label{amplitudes}
{\cal A}_+ &\!=\!& \frac{1}{2}\,\dot u^2\,(H_0)_{,\rho\rho}\,,\nonumber\\
{\cal A}_\times &\!=\!& \frac{1}{2}\,\dot u^2\frac{1}{\rho^2}\Big(\,\rho\, (H_0)_{,\rho\varphi}-(H_0)_{,\varphi}+2(\chi_{,u}-\omega \,\chi_{,\varphi})\Big) \,.
\end{eqnarray}
Recall that ${\omega=0}$ in the vacuum region outside the gyratonic source. The invariant equation of geodesic deviation (\ref{InvGeoDev}) in the interpretation frame, evaluated along the chosen timelike geodesic ${\gamma(\tau)}$, thus takes the explicit form
\begin{eqnarray}\label{geodetdev}
\ddot Z^{(1)} &\!=\!& 0\,,\nonumber\\
\ddot Z^{(2)} &\!=\!& {\cal A}_+\,Z^{(2)} +{\cal A}_\times\,Z^{(3)}\,, \\
\ddot Z^{(3)} &\!=\!& {\cal A}_\times\,Z^{(2)} -{\cal A}_+\,Z^{(3)}\,, \nonumber
\end{eqnarray}
where the functions ${{\cal A}_+}$ and ${{\cal A}_\times}$ obviously determine the two ``+'' and ``$\times$'' \emph{polarization amplitudes of the transverse gravitational waves propagating along} ${\bolde_{(1)}}$, respectively.

It is now straightforward to calculate the explicit forms of these wave amplitudes for the large family of exact solutions (\ref{function_H0}) when ${\chi=\chi(u)}$:
\begin{eqnarray}
{\cal A}_+ &\!=\!& \dot u^2\,\Bigg[\,\frac{\mu(u)}{\rho^2}\hspace{3mm}
+\frac{1}{2}\sum_{m=1}^\infty m(m+1)\,\frac{B_m(u)}{\rho^{m+2}}\,\cos\Big[m\Big(\varphi-\varphi_m(u)\Big)\Big] \nonumber\\
&&\hspace{21.3mm}
+\frac{1}{2}\sum_{m=2}^\infty m(m-1)\,A_m(u)\,\rho^{m-2}\cos\Big[m\Big(\varphi-\varphi'_m(u)\Big)\Big]\Bigg] \,,\label{amplitudes_for_H0+}\\
{\cal A}_\times &\!=\!& \dot u^2\,\Bigg[\,\frac{\chi_{,u}(u)}{\rho^2}
+\frac{1}{2}\sum_{m=1}^\infty m(m+1)\,\frac{B_m(u)}{\rho^{m+2}}\,\sin\Big[m\Big(\varphi-\varphi_m(u)\Big)\Big] \nonumber\\
&&\hspace{21.3mm}
-\frac{1}{2}\sum_{m=2}^\infty m(m-1)\,A_m(u)\,\rho^{m-2}\sin\Big[m\Big(\varphi-\varphi'_m(u)\Big)\Big]\Bigg] \,.\label{amplitudes_for_H0*}
\end{eqnarray}
The first terms $ \mu$ and $\chi_{,u}$ in each amplitude represent the
gravitational field of axially symmetric ``extended'' \emph{Aichelburg--Sexl
solution} \cite{AichelburgSexl:1971, Lessner:1986} with an ultrarelativistic
\emph{monopole} gyratonic source located along the axis ${\rho=0}$, which we
will describe in more detail in the next subsection~\ref{ASsolution}. The terms
$B_m$ correspond to asymptotically flat \emph{pp}-wave solutions with
\emph{multipole} gyratonic sources located along ${\rho=0}$
\cite{GriffithsPodolsky:1997,PodolskyGriffiths:1998,FrolovIsraelZelnikov:2005,
Podolsky:1998}. For example, the gyrating dipole source is given by the profile
function $B_1$ and $\chi_{,u}$. The gravitational field of the \emph{plane wave}
is described by $A_2$, in which case the wave amplitudes ${\cal A}_+$ and ${\cal
A}_\times$ are independent of the radial coordinate $\rho$.
\emph{Non-homogeneous} \emph{pp}-waves with directional curvature singularities
at ${\rho=\infty}$ (where ${{\cal A}_+ }$, ${{\cal A}_\times }$ diverge) are given by
higher-order terms $A_m$ with ${m=3,4,\ldots}$
\cite{PodolskyVesely:1998c,PodolskyVesely:1998d,PodolskyVesely:1999,
VeselyPodolsky:2000}. With ${\chi_{,u}\not=0}$ we obtain their gyrating versions
which manifest themselves in the amplitude $ {\cal A}_\times$ of geodesic
deviation (\ref{geodetdev}).

\subsection{The axially symmetric case}
\label{ASsolution}

The simplest vacuum \emph{pp}-wave solution with a gyratonic source is the axially symmetric one. It can be written in the form (\ref{general_metric}) with the metric functions (\ref{Jaftergauge1}) and (\ref{Haftergauge1}) independent of the angular coordinate $\varphi$. Since the only axially symmetric solution of (\ref{fieldeqhomogH}) is ${H=H_0= -2\,\mu(u)\, \log \rho}$, see (\ref{function_H0}), we are thus left with
\begin{equation}\label{axisym_functions_JH}
J(u) = \chi (u)\,, \qquad H(u,\rho) = -2\,\mu(u)\, \log \rho\,.
\end{equation}

The case ${\,J=0\,}$ represents the ``extended'' Aichelburg--Sexl solution because in the distributional limit when ${\mu(u)\to\delta(u)}$ we obtain the spacetime \cite{AichelburgSexl:1971} describing the specific impulsive gravitational wave generated by a nonrotating ultrarelativistic monopole point source located at ${\rho=0=u}$.

The case ${\,J\not=0\,}$ describes the \emph{Frolov--Fursaev gyraton} investigated in \cite{FrolovFursaev:2005}. There is a curvature singularity at ${\rho=0}$ whenever ${\mu\not=0}$. This can be immediately seen from the gravitational wave amplitudes (\ref{amplitudes_for_H0+}), (\ref{amplitudes_for_H0*}), which for (\ref{axisym_functions_JH}) simplify considerably to
\begin{equation}\label{amplitudesAS}
{\cal A}_+=\dot u^2\,\frac{\mu}{\rho^2}\,,\qquad
{\cal A}_\times=\dot u^2\,\frac{\chi_{,u}}{\rho^2}\,.
\end{equation}
Moreover, we conclude that the functions $\mu(u)$ and $\chi_{,u}(u)$ \emph{directly determine the} ``+'' \emph{and} ``$\times$'' \emph{polarization amplitudes of the gravitational waves}, respectively, as seen by the transverse deviations (\ref{geodetdev}) between the geodesic observers. Interestingly, \emph{both} these physically relevant functions in the metric coefficients $H$ and $J$ (determining the energy and angular momentum density of the null source) are thus directly observable by a detector of gravitational waves as the disctinct polarization states. For the non-spinning Aichelburg--Sexl source, the corresponding gravitational \emph{pp}-wave is purely ``+'' polarized since in the case ${\chi_{,u}=0}$ there is ${{\cal A}_\times=0}$.

Notice that for (\ref{axisym_functions_JH}) the mass-energy density (\ref{massdensity}) of the source can be explicitly evaluated. Using (\ref{fieldeqqq2}) we obtain
${\varrho = -\frac{1}{16\pi}\triangle\, H = \frac{1}{8\pi}\,\mu\, \triangle\, \log \rho= \frac{1}{4}\,\mu\,\delta^{(2)}}$, so that
\begin{equation}
{\cal M}(u) = {\textstyle\frac{1}{4\sqrt2}}\, \mu(u)
\,.
\end{equation}
The metric function $\mu(u)$ thus directly determines the mass-energy density ${\cal M}(u)$ of the gyratonic source. Since the source propagates with the speed of light, its rest mass must be zero which means that ${{\cal M}(u)= \frac{1}{4\sqrt2}\, \mu(u)}$ also equals to the \emph{density of momentum} while ${{\cal J}(u)=-\frac{1}{4}\,\chi(u)}$ is the \emph{density of angular momentum} of the gyraton.

\section{Geodesics}
\label{geodesics}

In this part we are going to analyze geodesics ${u\not=\,}$const.\ in gyratonic \emph{pp}-waves. Since all Christoffel symbols (\ref{Christ1}) of the form $\Gamma^u_{\alpha\beta}$ vanish we may choose $u$ as an \emph{affine parameter}. Setting $\dot u=1$ we arrive at
the following set of equations for the metric (\ref{general_metric}):
\begin{eqnarray}\label{geo-eq-polar}
&&\hspace{-7mm}
\ddot r-\frac{1}{2}H_{,u}+\frac{1}{2\rho^2} J(2J_{,u}-H_{,\varphi})-(H_{,\rho}-\frac{1}{\rho^2}JJ_{,\rho})\dot\rho
-H_{,\varphi}\dot\varphi +\frac{1}{\rho}(2J-\rho
J_{,\rho})\dot\rho\,\dot\varphi-J_{,\varphi}\dot\varphi^2=0 \,,\nonumber \\
&&\hspace{-7mm}
\ddot \rho-\rho\,\dot\varphi^2-
\frac{1}{2}H_{,\rho}-J_{,\rho}\,\dot\varphi=0\,,\\
&&\hspace{-7mm}
\ddot \varphi+\frac{2}{\rho}\,\dot\rho\,\dot\varphi
+\frac{1}{2\rho^2}\,(2J_{,u}-H_{,\varphi})+\frac{1}{\rho^2}\,J_{,\rho}\,
\dot\rho=0\,,\nonumber
\end{eqnarray}
where ${\dot{\,}=\frac{\d}{\d u}\,}$. The equation for $r$ is clearly decoupled and can be simply
integrated once the rest of the system has been solved. Applying the vacuum
field equations and the gauge leading to the form (\ref{Jaftergauge1}), (\ref{Haftergauge1}),
the transverese part simplifies to
\begin{equation}\label{geo-eq-polar-spatial}
\ddot\rho-\rho\,\dot\varphi^2-\frac{1}{2}H_{0,\rho}=0\,,\qquad
\ddot\varphi+\frac{2}{\rho}\,\dot\rho\,\dot\varphi+\frac{1}{\rho^2}\Big(\chi_{,u}
-\frac{1}{2}H_{0,\varphi}\Big)=0\,.
\end{equation}
In the case of axial symmetry, by (\ref{axisym_functions_JH}) the equations further reduce to
\begin{equation}\label{geo-eq-polar-spatial-1}
\ddot\rho-\rho\,\dot\varphi^2+\frac{\mu}{\rho}=0\,, \qquad
\ddot\varphi+\frac{2}{\rho}\,\dot\rho\,\dot\varphi+\frac{\chi_{,u}}{\rho^2}
=0\,.
\end{equation}
We observe that the mass-energy density proportional to $\mu(u)$ only occurs in the radial $\rho$-equation, while
the \emph{derivative} $\chi_{,u}$ of the angular momentum density proportional to $\chi(u)$ only appears
in the $\varphi$-equation. An analysis of (\ref{geo-eq-polar-spatial-1}) as in \cite{FrolovIsraelZelnikov:2005} shows
that positive energy qualitatively exerts an \emph{attractive force}
which leads to a focussing effect of the geodesics. On the other hand, angular momentum exerts a
\emph{rotational effect} on the geodesics. Note that we have already encountered
the analogous separation of effects in the geodesic deviation amplitudes (\ref{amplitudesAS}).
Due to the axial symmetry we also have a conserved quantity $\chi_0$ associated with the Killing vector $\partial_\varphi$
which enables us to rewrite the equations (\ref{geo-eq-polar-spatial-1}) as
\begin{equation}
\ddot\rho=-\frac{\mu(u)}{\rho}+\frac{[\,\chi_0-\chi(u)]^2}{\rho^3}\,, \qquad
\dot\varphi=\frac{\chi_0-\chi(u)}{\rho^2}\,.
\end{equation}
The first term in the equation for radial acceleration $\ddot\rho$ represents the focussing due to the positive energy density $\mu(u)$ of the source, while the second term is the nonlinear coupling to its angular momentum. The equation for the speed of rotation $\dot\varphi$ clearly involves the influence of the angular momentum density of the source ${{\cal J}(u)}$ proportional to $\chi(u)$, effectively adding to the conserved quantity $\chi_0$.

\section{Impulsive limit}
\label{impulsive limit}

In \cite{FrolovFursaev:2005,FrolovIsraelZelnikov:2005} impulsive versions of
gyratons have been introduced along with their extended versions and have been
used prominently in \cite{YoshinoZelnikovFrolov:2007}. Here we give a somewhat
broader discussion of possible impulsive limits in the class of gyratonic \emph{pp}-wave spacetimes.
We consider the vacuum line element (\ref{general_metric}) with
(\ref{Jaftergauge1}), (\ref{Haftergauge1}), that is
\begin{equation}\label{ro-metric}
\d s^2 = \d\rho^2+\rho^2\,\d\varphi^2-2\,\d u\,\d r +2\,\chi(u,\varphi)\,\d
u\,\d\varphi +H_0(u,\rho,\varphi)\,\d u^2 \,,
\end{equation}
where $\chi(u,\varphi)$ is an arbitrary function while $H_0(u,\rho,\varphi)$ is a solution of (\ref{Poisson}). There are now two distinct cases that have to be treated separately.

\subsection{The case ${\Sigma=0}$}
\label{Sigmais0}

With the constraint ${ \Sigma(u,\varphi)\equiv 2(\chi_{,u\varphi}-\omega\,\chi_{,\varphi\varphi})=0}$, the remaining vacuum field equation reduces to the Laplace equation ${\triangle\, H_0 = 0}$. In the natural global gauge ${\omega=0}$, the constraint implies that the function $\chi_{,u}$ is independent of $\varphi$, so that ${\chi(u,\varphi)=\chi(u)+\Phi(\varphi)}$. In such a case we see that \emph{the field equations put no restriction} on the $u$-dependence of $H_0$ and $\chi$. In analogy with the usual (non-gyratonic) class of sandwich and impulsive \emph{pp}-waves in Minkowski space
(and corresponding models with a nonvanishing cosmological constant \cite{Podolsky:1998, Podolsky:2002b}) we now consider the metric functions of the form
\begin{equation}\label{ro-profiles}
H_0(u,\rho,\varphi)= \tilde H_0(\rho,\varphi)\,\chi_H(u)\,,\qquad
\chi (u,\varphi) = J(u,\varphi) = \tilde \chi \,\chi_J(u)+\Phi(\varphi)\,,
\end{equation}
where $\tilde \chi$ is a constant and we call the functions $\chi_H(u)$ and $\chi_J(u)$ \emph{profile functions} of the energy and angular momentum densities, respectively. Here we assume $\int\!\chi_H(u)\,\d u=1=\int\!\chi_J(u)\,\d u$ but otherwise these functions are completely arbitrary. In particular, these profiles can be \emph{choosen independently of each other} --- we have a complete separation of $\chi_H(u)$ and $\chi_J(u)$. From (\ref{Riemgen}) it then follows that
\begin{eqnarray}\label{Riemgenprofile}
&&
R_{u\rho u\rho}=-\frac{1}{2}(\tilde H_0)_{,\rho\rho}\,\chi_H(u) \,,
\nonumber\\
&&
R_{u\rho u\varphi}=-\frac{1}{2\rho}\Big(\rho\, (\tilde H_0)_{,\rho\varphi}-(\tilde H_0)_{,\varphi} \Big)\,\chi_H(u)
-\frac{\tilde\chi}{\rho}\,{\chi_J}_{,u}(u) \,,
\\
&&
R_{u\varphi u\varphi}=-\frac{1}{2}\Big((\tilde H_0)_{,\varphi\varphi}+\rho (\tilde H_0)_{,\rho}\Big)\,\chi_H(u) \,,\nonumber
\end{eqnarray}
which in the axisymmetric case (\ref{axisym_functions_JH}) with ${\chi_H(u)=\mu(u)}$ and ${\chi(u)=\tilde\chi\,\chi_J(u)}$ reduce to
\begin{equation}
R_{u\rho u\rho}=-\frac{1}{\rho^2} \,\mu(u)\,,\quad
R_{u\rho u\varphi}=-\frac{1}{\rho}\,\chi_{,u}(u)\,,\quad
R_{u\varphi u\varphi}=\mu(u)\,. \label{Sigma0curvatures}
\end{equation}
We thus observe, that the energy profile $\chi_H(u)$ explicitly shows up in the curvature, and so
does the angular momentum profile $\chi_J(u)$ in term of its \emph{derivative} ${\chi_J}_{,u}(u)$. This is in accordance with (\ref{amplitudes}) where the energy profile determines the amplitude ${\cal A}_+$, while the angular momentum profile determines the amplitude ${\cal A}_\times$, which then also contains the derivative of the profile. The same effect becomes visible in the only nonvanishing field scalar $\Psi_4$ (\ref{Psi4vac}), which in the axially symmetric case takes the form
\begin{equation}
\Psi_4=-\frac{1}{\rho^2} \,\mu(u)
+\frac{\iu}{\rho^2}\,\chi_{,u}(u) \,.\label{Psi4vacaxi}
\end{equation}

All this suggests that --- in accordance with the usual definition of impulsive waves ---
one may take the \emph{energy profile} $\chi_H(u)$ to be $\delta$-shaped (the Dirac $\delta$ distribution) but one should rather confine oneself with a \emph{box-like profile for the angular momentum} $\chi_J(u)$. Indeed, a $\delta$-shaped angular momentum profile would introduce a $\delta'$-term in the curvature, the amplitude ${\cal A}_\times$ and the filed scalar $\Psi_4$. This seems to be less physical.

\begin{figure}[t]
\begin{center}\includegraphics[scale=0.55]{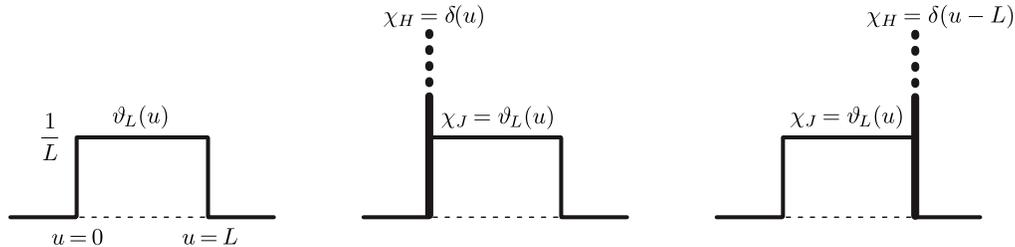}%
\caption{Schematic representation of possible profiles $\chi_H(u)$ for energy and $\chi_J(u)$ for angular momentum densities: a box-shaped profile $\vartheta_L(u)$ for both $\chi_H$ and $\chi_J$ (left); the choice ${\chi_H=\delta(u)}$ with ${\chi_J=\vartheta_L(u)}$
(middle); alternative choice ${\chi_H=\delta(u-L)}$ with ${\chi_J=\vartheta_L(u)}$ (right).}
\label{fig2}
\end{center}
\end{figure}

Indeed in \cite{FrolovFursaev:2005} such a box-shaped profile for \emph{both} energy and angular momentum densities was considered, i.e.,
${\chi_H(u)=\chi_J(u)=\vartheta_L(u)}$ where
\begin{equation}\label{box}
\vartheta_L(u)\equiv\frac{1}{L}\Big(\Theta(u)-\Theta(u-L)\Big) \,,
\end{equation}
in which ${L>0}$ is the ``length'' of the profile and $\Theta$ denotes the Heaviside step function, see the left part of figure~\ref{fig2}.

In \cite{YoshinoZelnikovFrolov:2007} the energy profile was changed to be impulsive, more precisely ${\chi_H(u)=\delta(u)}$, as shown in the middle part of figure~\ref{fig2}, or alternatively ${\chi_H(u)=\delta(u-L)}$, see the right part of figure~\ref{fig2}. This is possible since
\emph{arbitrary} and \emph{independent} profile functions $\chi_H(u)$ and $\chi_J(u)$ can be prescribed in the ${\Sigma=0}$ case.

\subsection{The case ${\Sigma\not=0}$}
\label{Sigmaisnot0}

\begin{figure}[t]
\begin{center}\includegraphics[scale=0.55]{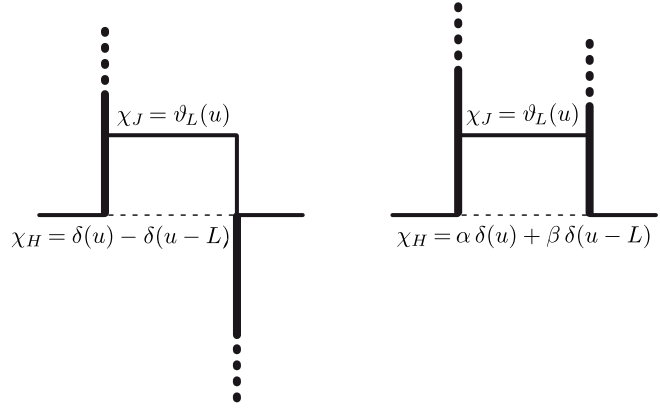}%
\caption{Possible profiles $\chi_H(u)$ for energy density in the case of a general $\Sigma$ and the box-shaped angular momentum profile ${\chi_J(u)=\vartheta_L(u)}$: the natural choice ${\chi_H=\delta(u)- \delta(u-L)}$ (left); more general choice ${\chi_H=\alpha\, \delta(u)+\beta\, \delta(u-L)}$ (right).}
\label{fig3}
\end{center}
\end{figure}

If the metric function ${J=\chi(u,\varphi)}$ nontrivially depends on the angular coordinate $\varphi$ then ${\Sigma\not=0}$ and \emph{the field equations naturally lead to a coupling of the profiles}. More precisely, the vacuum equation (\ref{Poisson}) with ${\omega=0}$
then reduces to
\begin{equation}\label{eqtosolve}
\triangle H_0(u,\rho,\varphi)=\frac{2}{\rho^2}\ \chi_{,u\varphi}(u,\varphi),
\end{equation}
and the splitting generalizing (\ref{ro-profiles}) to admit ${\tilde \chi= \tilde \chi(\varphi)}$,
\begin{equation}\label{ro-profilessosig}
H_0(u,\rho,\varphi)= \tilde H_0(\rho,\varphi)\,\chi_H(u)\,,\qquad
\chi (u,\varphi) = \tilde \chi(\varphi) \,\chi_J(u)+\Phi(\varphi)\,,
\end{equation}
gives
\begin{equation}\label{Sigmanot0}
\triangle \tilde H_0(\rho,\varphi)\,\chi_H(u)=\frac{2}{\rho^2}\ \tilde \chi_{,\varphi}(\varphi) \,{\chi_J}_{,u}(u)\,.
\end{equation}
If the supports of $\chi_H(u)$ and $\chi_J(u)$ are disjoint, then separately ${\triangle \tilde H_0=0}$ and ${\tilde \chi_{,\varphi}=0}$, implying ${\Sigma=0}$. Thus, for solutions with ${\Sigma\not=0}$, \emph{both the supports must agree}, in which case the box profile (\ref{box})
in the angular momentum density, ${\chi_J(u)=\vartheta_L(u)}$, \emph{naturally} leads to two Dirac deltas in the energy density, namely ${\chi_H(u)=\delta(u)- \delta(u-L)}$ (to satisfy the relation ${\chi_H(u)\propto{\chi_J}_{,u}(u)}$ in (\ref{Sigmanot0})), see the left part of figure~\ref{fig3}. Assuming (\ref{ro-profilessosig}) we thus obtain an impulse in energy on each edge of the box, one positive and the other negative. Hence, the energy distribution in such a gyratonic source would have a dipole character. Moreover, the corresponding spatial Poisson equation
\begin{equation}
\triangle \tilde H_0(\rho,\varphi)=\frac{2}{L}\, \frac{\tilde \chi_{,\varphi}(\varphi)}{\rho^2}\,,
\end{equation}
also has to be solved.

The drawback of the particular solution ${\chi_H(u)=\delta(u)- \delta(u-L)}$ is that it involves a \emph{negative} energy distribution located at ${u=L}$. However, physically more relevant solutions can be constructed by \emph{superimposing} this particular solution with specific \emph{homogeneous} solutions proportional to a positive profile $\delta(u-L)$ (and also possibly $\delta(u)$), effectively leading to a general energy density profile function ${\chi_H=\alpha\, \delta(u)+\beta\, \delta(u-L)}$ with \emph{positive} parameters $\alpha$ and $\beta$, see the right part of figure~\ref{fig3}.

In fact, as suggested by the referee, more general solution of the linear equation (\ref{eqtosolve}) takes the form ${H_0=H_0^{(h)}(u,\rho,\varphi)+H_0^{(p)}(u,\rho,\varphi)}$, where ${\triangle H_0^{(h)}=0}$ and ${\triangle H_0^{(p)}=2\rho^{-2}\chi_{,u\varphi}\,}$ with the additional requirement
\begin{equation}\label{constrnonhomog}
\oint H_0^{(p)}(u,\rho,\varphi)\,\d \varphi=0\,.
\end{equation}
The corresponding energy profiles ${\chi_H^{(h)}(u)}$ and ${\chi_H^{(p)}(u)}$ are independent. We can choose ${\chi_H^{(h)}(u)}$ in such a way that the energy is positive (the case ${\Sigma=0}$). Due to the additional constraint (\ref{constrnonhomog}) the energy of the \emph{complete} solution ${H_0=H_0^{(h)}+H_0^{(p)}}$ remains positive because the contribution from ${H_0^{(p)}}$ becomes zero after the $\varphi$-integration.

\section{Completeness of geodesics in the impulsive limit}
\label{deodinimpuls}

Finally, we analyze the geodesic equations in the impulsive limit of the ${\Sigma=0}$ case, i.e., the
models extending those of \cite{YoshinoZelnikovFrolov:2007},
also discussed in subsection~\ref{Sigmais0}. Specifically, we consider the metric (\ref{ro-metric}), (\ref{ro-profiles}) with
\begin{equation}
\chi_H(u)=\alpha\, \delta(u)+\beta\, \delta(u-L)\,, \qquad
\chi_J(u)=\vartheta_L(u)\,,
\end{equation}
where the box-profile ${\vartheta_L(u)}$ is defined in (\ref{box}) and $\alpha$, $\beta$ are \emph{arbitrary} real parameters, indicated in the right part of figure~\ref{fig3}. They enable us to set the respective impulsive components to any values (including turning them off). Using (\ref{geo-eq-polar-spatial}) we find that the spatial part of the geodesic equations takes the form
\begin{eqnarray}
\ddot\rho-\rho\,\dot\varphi^2-\frac{1}{2}\,\tilde H_{0,\rho}(\rho,\varphi)
\Big(\alpha\, \delta(u)+\beta\,\delta(u-L)\Big)=0\,,&& \\
\ddot\varphi+\frac{2}{\rho}\,\dot\rho\,\dot\varphi+\frac{1}{\rho^2}\,\frac{
\tilde \chi}{L}\Big(\delta(u)+\delta(u-L)\Big)-\frac{1}{2\rho^2}\tilde H_{0,\varphi}(\rho,\varphi)
\Big(\alpha\, \delta(u)+\beta\,\delta(u-L)\Big)
=0\,.&&
\end{eqnarray}
These equations have a form similar to the geodesic equations in impulsive
NP-waves, which have recently been analyzed rigorously in \cite{SaemannSteinbauer:2012, SaemannSteinbauer:2014}.
This is best seen by looking at the geodesic equations in Cartesian coordinates $x^i$ of the metric (\ref{pp_metric}), which using (\ref{Christ2}) are
\begin{eqnarray}\label{geo-eq}
&&{\textstyle \ddot x^i + \delta^{ij}(a_{j,u}-\frac{1}{2}H_{,j}) +
\delta^{ik}(a_{k,j}-a_{j,k})\,\dot x^j = 0}\,.
\end{eqnarray}
In view of (\ref{transf}), in the present case they are of the form
\begin{equation}
\ddot x(u)+G_1(x)\,\delta(u)+G_2(x)\,\delta(u-L)+G_3(x,\dot x)\,\vartheta_L(u)=0\,,
\end{equation}
where for simplicity we have collected all dependencies on the spatial variables ${x=(x^2,x^3)}$,
spanning the transverse plane, in the functions ${G_1, G_2, G_3}$.
In particular, assuming these functions to be \emph{smooth}, the technical result
derived in \cite[Lem.\ A.2]{SaemannSteinbauer:2012} (with minor modifications, namely
replacing $F_1(x_\epsilon,\dot x_\epsilon)$ by $F_1(x_\epsilon,\dot x_\epsilon)\Theta_\epsilon$)
also applies in the present situation and we obtain a \emph{completeness result}. More
precisely, if one regularizes the profile functions replacing $\delta$ by a
standard mollifier ${\delta_\epsilon(x)\equiv
(1/\epsilon)\,\phi(x/\epsilon)}$ ($\phi$ a smooth function supported in
$[-1,1]$ with unit integral) and replacing $\Theta$
by ${\Theta_\epsilon(x)\equiv\int_{-1}^x\phi_\epsilon(t)\,\d t}$ one obtains
the following completeness statement: \emph{For any geodesic starting long before the
shock, say at $u=-1$, there exists an $\epsilon_0$ such that it passes the
shock region ${u\in[0,L]}$ provided ${\epsilon\leq\epsilon_0}$}. The geodesics will
be straight lines up to ${u=0}$ where they will be refracted with their
$r$-component suffering an additional jump. During the support of the angular momentum
profile ${u\in [0,L]}$ an angular motion is excerted. Another break occurs at
${u=L}$ and after that the geodesics return to be straight lines.

In the more realistic case where the ${G_1, G_2, G_3}$ are non-smooth on the axis ${\rho=0}$ (including the Aichelburg--Sexl solution)
the result still applies with the exception of those geodesics which are
directly heading into the curvature singularity at ${u=0}$, ${\rho=0}$ and those which are
refracted to directly hit the singularity at ${u=L}$, ${\rho=0}$. A more detailed
analysis of the geodesic motion in these and more general models is subject to
current research.

\section{Conclusions}

We have studied the complete family of \emph{pp}-waves with flat wavefronts in
Einstein's general relativity. Thereby we have kept all terms in the original
Brinkmann form of the metric~(\ref{pp_metric}), in particular, the off-diagonal
ones ${a_i(u,x^j)\,\d u\,\d x^i}$.
In almost all prior investigations of this famous family
of exact spacetimes the functions $a_2$ and~$a_3$ have been ignored because (in
any vacuum region) it is always possible locally to set ${a_i=0}$ by a suitable
gauge. However, it was only Bonnor in 1970 and independently Frolov and his
collaborators in 2005 who pointed out the physical significance of these
off-diagonal terms. In general, they cannot be removed globally, and in such a
case they encode angular momentum. For a localized matter distribution the
spacetime metric can be interpreted as the gravitational field of a
spinning source moving at the speed of light, called \emph{gyraton}
\cite{FrolovFursaev:2005,FrolovIsraelZelnikov:2005}.
Here we have significantly extended and complemented the previous studies
\cite{Bonnor:1970b,FrolovFursaev:2005,
FrolovIsraelZelnikov:2005,YoshinoZelnikovFrolov:2007} explicitly investigating
various mathematical, geometrical and physical properties of these spacetimes.

We have fully integrated the vacuum Einstein equations in section~\ref{integrating},
yielding the general metric functions $J$ and $H$, see (\ref{explicit_metric_function_J})
and (\ref{explicit_metric_function_H}). Then we have used the complete gauge
freedom\footnote{Let us remark that our approach differs from those in
\cite{Bonnor:1970b,FrolovFursaev:2005,FrolovIsraelZelnikov:2005,
YoshinoZelnikovFrolov:2007}. There the gauge freedom was
considered \emph{before} solving the field equations which somehow obscures the
significance and meaning of the integration functions $\omega$ and $\Sigma$.}
to understand the geometrical and physical meaning of all the integration
functions $\omega(u)$, $\chi(u,\varphi)$ and $H_0(u,\rho,\varphi)$.
We showed that the function $\omega(u)$ represents angular velocity of a
rigid rotation of the whole spacetime, which can always be set to zero in the
vacuum region by using (\ref{gauge1}). With ${\omega=0}$ it is then possible to
employ the second gauge (\ref{gauge2cart}) to set the function
${J=\chi(u,\varphi)}$ to zero since the integrability conditions
${\Boldomega =\frac{1}{2}\,J_{,\rho}\,\boldd \rho\wedge \boldd \varphi=0}$
are satisfied, see (\ref{geomintegrabcondit2}), (\ref{J0}). However, this can be done \emph{only
locally}. In fact, the key $2$-form ${\Boldomega}$ is given by ${\Boldomega = \frac{1}{2}\,\boldd
\bolda}$, where ${ \bolda\equiv a_i\,\boldd x^i }$, cf. (\ref{omega2formdef}),
and in general the closed $1$-form $\bolda$ need not be \emph{globally exact}.
This happens, in particular, if a gyratonic source is located along the axis
${\rho=0}$, so that the external vacuum region is not contractible. Then it is
\emph{necessary} to keep the off-diagonal metric terms even in the vacuum
region.

Moreover, as shown in \cite{Bonnor:1970b, FrolovFursaev:2005} and
subsection~\ref{energyANDmomentum}, the function $H$ is related to the mass-energy
density ${\cal M}(u)$ while $J$ encodes the angular momentum density ${\cal
J}(u)$ of the gyratonic source via the gauge-independent contour integral
${{\cal J}(u) = -\frac{1}{8\pi}\oint_C \bolda =
-\frac{1}{8\pi}\oint_C J\,\boldd \varphi}$, 
cf. (\ref{integralStokesfinal}), (\ref{integral}). This yields a clear physical
interpretation of the metric functions.

To analyze further aspects of the gyratonic sources represented by the metric
functions~$a_i$ that have not been studied before, we introduced
in subsection~\ref{interpretationframe} a natural orthonormal frame
for any geodesics observer and its associated null frame~(\ref{null_frame_pp}). After
studying their gauge freedom, in subsection~\ref{rotation}, we analyzed the
rotational dragging effect of parallelly propagated frames caused by the
spinning gyratonic matter. Specifically, we proved that the angular velocity of the
spatial rotation of such frames is given by
${ {\Omega}^{\,2}_{\ 3} = {\dot u}\, \omega(u,\rho,\varphi)}$, where ${\omega =
\frac{1}{2}J_{,\rho}/\rho =\frac{1}{2}(a_{3,2}-a_{2,3})}$
is the only nontrivial component of the key 2-form $\Boldomega$. Therefore, in the external
vacuum region there is no such dragging while inside the gyratonic source, where
${\Boldomega\not=0}$, we have rotation of frames, see figure~\ref{fig1}.

In subsection~\ref{NPscalars} we analyzed the Newman--Penrose scalars
(\ref{Psi4})--(\ref{Phi12}). The spacetime inside the
source is of Petrov type~III, with the gyrating matter component
$\Phi_{12}$ coupled to the gravitational component $\Psi_3$ so that they (do
not) vanish simultaneously. The gravitational field in the exterior vacuum
region is thus of type~N with $\Psi_4$ given by (\ref{Psi4vac}).

As a further new application of the interpretation frame, in section~\ref{frame}
we investigated the invariant form of the geodesic deviation. This allows to clearly separate
the longitudinal direction, in which the gravitational wave propagates, and the
transverse 2-space where its effect on relative motion of test particles is
observed (\ref{geodetdev}). We explicitly evaluated the two polarization
amplitudes ${{\cal A}_+}$ and ${{\cal A}_\times}$, see (\ref{amplitudes}), which
turn out to be proportional to the real and the imaginary part of $\Psi_4$,
respectively. In the axisymmetric case with ${J=\chi(u)}$,
${H=-2\,\mu(u)\,\log \rho}$ (the Frolov--Fursaev gyraton constructed from the
Aichelburg--Sexl solution), which we have studied in subsection~\ref{ASsolution},
the amplitudes are given by ${{\cal A}_+=\mu\,\dot u^2\rho^{-2}}$ and ${{\cal
A}_\times=\chi_{,u}\,\dot u^2\rho^{-2}}$.
Interestingly, both physical
profile functions $\mu(u)$ and $\chi(u)$ of $H$ and $J$ (determining the energy and angular momentum densities) are
thus observable by a detector of gravitational waves as distinct ``+'' and
``$\times$'' polarization states, with the wave being
purely ``+'' polarized in the absence of a gyraton.
It follows from (\ref{amplitudes_for_H0+}), (\ref{amplitudes_for_H0*}) that a
similar behavior also occurs for non-axisymmetric gyratons with multipole
sources.

It is important to observe that it is the \emph{derivative}
$\chi_{,u}$ of the metric function ${J=\chi(u, \varphi)}$ which
appears in the curvature $\Psi_4$ and in the wave amplitude ${\cal A}_\times$.
Moreover, $\chi_{,u}$ directly influences the behavior of geodesics, studied in
section~\ref{geodesics}, namely the \emph{axial acceleration} $\ddot\varphi$. On
the other hand, the mass-energy density of the gyraton encoded in
${H=H_0(u,\rho,\varphi)}$ determines the \emph{radial acceleration} $\ddot\rho$,
causing a focusing of the geodesics.

We have studied \emph{impulsive limits} of gyratonic
\emph{pp}-waves in section~\ref{impulsive limit}. There we
emphasized that (unlike in previous works) it is necessary to distinguish the cases $\Sigma=0$ and
$\Sigma\not=0$, when solving the Poisson equation (\ref{Poisson}).
If ${\Sigma=0}$ it reduces to the Laplace equation
and the field equations put \emph{no restriction} on the
$u$-dependence of $H_0$ and $J$. Hence the corresponding profiles $\chi_H(u)$ and
$\chi_J(u)$ of the energy and the angular momentum densities can be chosen
independently of each other. Since by (\ref{Riemgenprofile}) the
curvature is proportional to $\chi_H$ and ${\chi_J}_{,u}$, it is natural to consider
impulsive waves
by setting the profile $\chi_H(u)$ to be proportional to the \emph{Dirac} $\delta$ but
using a \emph{box-like profile} $\vartheta_L(u)$
for $\chi_J(u)$, see (\ref{box}) and figure~\ref{fig2}.

On the other hand, when $\Sigma\not=0$ the \emph{supports} of $\chi_H(u)$ and
${\chi_J}_{,u}(u)$ \emph{must coincide}. In particular, the box $\vartheta_L(u)$ in
the angular momentum density profile $\chi_J(u)$ naturally leads to two
Dirac deltas in the energy density, ${\chi_H=\delta(u)- \delta(u-L)}$. These two
(opposite) impulses on each edge of the box are shown in the left part of
figure~\ref{fig3}. Physically more relevant solutions can be constructed by superimposing such a particular solution with specific \emph{homogeneous} solutions. This effectively leads to a general energy density profile ${\chi_H=\alpha\, \delta(u)+\beta\, \delta(u-L)}$ with \emph{positive} parameters $\alpha$ and $\beta$, see the right part of figure~\ref{fig3}.

In the final section~\ref{deodinimpuls} we analyzed the geodesic equations in
the impulsive limit. We considered the ${\Sigma=0}$ case and the generic
profiles ${\chi_H=\alpha\, \delta(u)+\beta\, \delta(u-L)}$ and
${\chi_J=\vartheta_L(u)}$ where $\alpha$, $\beta$ are arbitrary constants. We showed that any geodesic starting long
before such a wave \emph{passes the shock region} ${u\in[0,L]}$, considering
regularizations of the Dirac deltas and the box by standard smooth mollifiers.
In other words, we proved geodesic completeness of impulsive gyratonic
\emph{pp}-wave spacetimes.

\section*{Acknowledgments}

We thank Pavel Krtou\v{s} for very useful comments on the manuscript and
Clemens S\"amann for his contributions in our joint discussions.
JP and R\v{S} were supported by the grant GA\v{C}R~P203/12/0118, project
UNCE~204020/2012 and the grant 7AMB13AT003 of the Scientific and Technological
Co-operation Programme Austria--Czech Republic. RS acknowledges the support of
its Austrian counterpart, OEAD's WTZ grant CZ15/2013 and of FWF grant P25326.

\end{document}